\documentclass[a4paper,12pt]{article}
\pdfoutput=1
\usepackage{graphicx}
\usepackage{mathtools}
\usepackage{amssymb}
\usepackage{amsfonts}
\usepackage[footnotesize]{caption}
\usepackage[hang]{footmisc}
\usepackage[font=scriptsize]{subcaption}
\usepackage{braket}
\usepackage{cite}
\usepackage{hyperref}
\usepackage{url}
\usepackage{multirow}
\usepackage{relsize}
\usepackage{fullpage}
\usepackage{makecell}
\usepackage{blkarray}
\usepackage{fullpage}
\usepackage{grffile}
\usepackage{float}
\usepackage[labelfont=bf,textfont=md]{caption}
\usepackage{makecell}
\usepackage[svgnames]{xcolor}
\usepackage{tikz}
\usepackage{bm}
\usepackage{gensymb}
\usepackage{amsmath}
\usepackage{booktabs}

\usetikzlibrary{shapes,arrows}

\tikzstyle{decision} = [diamond, draw, fill=blue!20, 
text width=4.5em, text badly centered, node distance=3cm, inner sep=0pt]
\tikzstyle{block} = [rectangle, draw, fill=Olive!15, 
text width=6em, text centered, rounded corners, minimum height=2em]
\tikzstyle{block0} = [rectangle, draw, fill=SteelBlue!20, 
text width=5em, text centered, rounded corners, minimum height=2em]
\tikzstyle{line} = [draw, -latex']
\tikzstyle{cloud1} = [draw, ellipse,fill=Linen, node distance=3cm,
minimum height=2em]
\tikzstyle{cloud2} = [draw, ellipse,fill=red!40, node distance=3cm,
minimum height=2em]
\tikzstyle{io} = [trapezium, trapezium left angle=70, trapezium right angle=110, minimum width=3cm, minimum height=1cm, text centered, draw=black, fill=DarkRed!25]

\newcommand{\newc}{\newcommand}
\newc{\be}{\begin{equation}}
\newc{\ee}{\end{equation}}
\newc{\bea}{\begin{eqnarray}}
\newc{\eea}{\end{eqnarray}}
\newc{\simlt}{~\mbox{\smaller\(\lesssim\)}~}
\newc{\simgt}{~\mbox{\smaller\(\gtrsim\)}~}

\newcommand{\pmatr}[1]{\begin{pmatrix} #1 \end{pmatrix}}

\newcommand{\phiatm}{\phi_{\mathrm{atm}}}
\newcommand{\phisol}{\phi_{\mathrm{sol}}}

\newcommand{\dof}{N_{d.o.f.}}

\begin{document}

\begin{titlepage}
\begin{center}
{\bf\Large
\boldmath{Fitting high-energy Littlest Seesaw parameters using low-energy neutrino data and leptogenesis
}
} \\[12mm]
Stephen~F.~King$^{\star}$%
\footnote{E-mail: \texttt{King@soton.ac.uk}},
Susana Molina Sedgwick$^{\star, \ddagger}$%
\footnote{E-mail: \texttt{S.Molina-Sedgwick@soton.ac.uk}},
Samuel~J.~Rowley$^{\star}$%
\footnote{E-mail: \texttt{S.Rowley@soton.ac.uk}},
\\[-2mm]
\end{center}
\vspace*{0.50cm}
\centerline{$^{\star}$ \it
School of Physics and Astronomy, University of Southampton,}
\centerline{\it
SO17 1BJ Southampton, United Kingdom }
\vspace*{0.20cm}
\centerline{$^{\ddagger}$ \it
Particle Physics Research Centre, Queen Mary University of London,}
\centerline{\it
E1 4NS London, United Kingdom }

\vspace*{1.20cm}

\begin{abstract}
{\noindent
We show that the four high-energy Littlest Seesaw parameters in the flavour basis,
namely two real Yukawa couplings plus the two right-handed neutrino masses, can be determined by an excellent fit to the seven currently constrained observables of low-energy neutrino data and leptogenesis. 
Taking into account renormalisation group corrections, we estimate $\chi^2 \simeq 1.5-2.6$ for the three d.o.f., depending on the high-energy scale and the type of non-supersymmetric Littlest Seesaw model.
We extract allowed ranges of neutrino parameters from our fit data, including the approximate mu-tau
symmetric predictions $\theta_{23}=45^o\pm 1^o$
and $\delta = -90^o \pm 5^o $, which, together with a
normal mass ordering with $m_1=0$, will enable Littlest Seesaw models to be tested in future neutrino experiments.}
\end{abstract}
\end{titlepage}

\section{Introduction}
\label{intro}
Neutrino oscillation experiments have provided the first solid evidence for \textbf{new physics} beyond the \textbf{Standard Model} (BSM) in the form of neutrino mass and mixing
\cite{nobel}. However, the theoretical origin of neutrino mass generation and lepton flavour mixing remains unknown~\cite{XZbook,King:2013eh}. 
In addition, significant uncertainties remain in the mixing parameters; for example,
the octant of the atmospheric angle is not yet determined, and its precise value is unknown. While T2K has long preferred a close to maximal atmospheric mixing angle~\cite{Abe:2017uxa}, NO$\nu$A originally excluded maximal mixing at $2.6\sigma$ CL~\cite{Adamson:2017qqn}, though the latest analysis with more data is now consistent with maximal mixing \cite{NOvA_new,deSalas:2017kay,Esteban:2016qun}.
Furthermore, the CP violating Dirac phase relevant for neutrino oscillations has not been directly measured.
The leading candidate for a theoretical explanation of neutrino mass and mixing remains 
the original \textbf{type I seesaw mechanism}~\cite{Minkowski:1977sc, Yanagida:1979ss, Gell-Mann:1979ss, Glashow:1979ss, Mohapatra:1979ia,Schechter:1980gr} involving right-handed neutrinos.

Although the type I seesaw mechanism
provides a very attractive mechanism for understanding the smallness of neutrino masses, it generally
involves a large number of free parameters and is therefore unpredictive.
In the flavour basis, where the charged lepton mass matrix and the right-handed neutrino mass matrix are both diagonal,
there are typically a larger number of undetermined Yukawa couplings and phases than low energy observables, with the precise number depending on the number of right-handed neutrinos~\cite{Schechter:1980gr}. 
This means that it is not possible to uniquely determine  the high-energy seesaw parameters from low-energy neutrino data
for the general type I seesaw model, making the seesaw model unpredictive. If the right-handed neutrino masses are above the TeV scale, as generally expected by leptogenesis, then it is also difficult to test directly. This motivates the study of minimal seesaw models involving fewer input parameters. If the number of input parameters is less than the number of observables, then such seesaw models become predictive.

One approach to reducing the number of seesaw parameters is to consider a
\textbf{minimal version} involving either \textbf{one}~\cite{King:1998jw}
or \textbf{two right-handed neutrinos} (2RHN)~\cite{King:1999mb,King:2002nf}.
The original 2RHN model - with one texture zero and sequential dominance (SD)
of the right-handed neutrinos - predicted a normal ordering (NO) with 
the lightest neutrino being massless, $m_1=0$.
Subsequently, a 2RHN model was proposed~\cite{Frampton:2002qc}
with two texture zeros in the Dirac neutrino mass matrix, 
consistent with cosmological leptogenesis~\cite{Fukugita:1986hr,Guo:2003cc, Ibarra:2003up, Mei:2003gn, Guo:2006qa, Antusch:2011nz,
Harigaya:2012bw, Zhang:2015tea,Branco:2005jr}. 
However, the 2RHN model with two texture zeros is only compatible with an inverted ordering (IO) of neutrino masses~\cite{Harigaya:2012bw, Zhang:2015tea}.
Thus, present data favours the 2RHN model with one texture zero as originally proposed~\cite{King:1999mb,King:2002nf}. 

In order to increase predictivity further, a constrained form of the Yukawa matrix with columns of magnitude 
proportional to $(0,1,1)$ and $(1,1,1)$ in the flavour basis - called \textbf{constrained sequential dominance} (CSD) -
was later proposed, 
which led to \textbf{tri-bimaximal mixing} \cite{King:2005bj}.
Following the measurement of a non-zero reactor angle, generalised constrained forms of Yukawa matrix
in the flavour basis were proposed - with columns of magnitude 
proportional to $(0,1,1)$ and $(1,n,n-2)$ or $(1,n-2,n)$, where $n$ is an integer -
called CSD(n),
again leading to other highly predictive schemes
~\cite{Antusch:2011ic,King:2013iva,King:2013xba,King:2013hoa,King:2014iia,
Bjorkeroth:2014vha,Bjorkeroth:2015ora,Bjorkeroth:2015tsa,King:2015dvf,King:2016yvg,
Ballett:2016yod,Ding:2018fyz,Chianese:2018dsz,King:2018kka}.	 
The most successful of these was CSD(3), with 
Yukawa columns in the flavour basis proportional to $(0,1,1)$ and $(1,3,1)$ or $(1,1,3)$,
and with real constants of proportionality $a$ and $b$, respectively,
together with a fixed relative phase between these columns of $\mp \pi/3$~\cite{King:2013iva}.

The successful CSD(3) scheme~\cite{King:2013iva} 
was later renamed as the \textbf{Littlest Seesaw} (LS) model~\cite{King:2015dvf},
to emphasise its status as the most minimal seesaw model which can explain current data 
with the smallest number of parameters, namely two right-handed neutrino masses plus the
real coefficients $a,b$ of the two column vectors comprising the Yukawa matrix, in the flavour basis.
It was later shown that the LS model could be obtained from an $S_4$ family symmetry together with 
other discrete $Z_N$ symmetries~\cite{King:2016yvg,Ding:2018fyz}.
Recently it has been shown that the LS model has an approximate accidental mu-tau reflection symmetry, which accounts for 
its approximate predictions of maximal atmospheric mixing and maximal CP violation~\cite{King:2018kka}.
The prospects of testing such LS predictions at future experiments has also been studied~\cite{Ballett:2016yod}
in the absence of renormalisation group corrections.

\textbf{Renormalisation group} (RG) corrections to the LS model have been considered in~\cite{King:2016yef}, 
including a comprehensive 
$\chi^2$ analysis of the low energy masses and mixing angles, 
in the presence of RG corrections,
for various right-handed neutrino masses and mass orderings, both with and without supersymmetry
\cite{Geib:2017bsw}. In particular, it was shown that the heavier RHN mass strongly affects the RG corrections.
However, \textbf{leptogenesis} was not included in either of these analyses.
Since leptogenesis is mainly controlled by the lighter RHN in the LS model~\cite{Bjorkeroth:2015tsa}, 
we are motivated in this paper to include leptogenesis in the \textbf{global fit}, in order to fix both RHN masses.
This enables both the high-energy RHN masses and the Yukawa coupling constants $a,b$
to be fixed by low-energy neutrino data and leptogenesis for the first time in any seesaw model.

In this paper we show that the four \textbf{high-energy LS parameters} in the flavour basis - two real
\textbf{Yukawa couplings} $a,b$
plus the two \textbf{right-handed neutrino masses} - can be determined by an excellent fit to the seven currently constrained observables of low-energy neutrino data and leptogenesis. 
Although there are in effect ten observables (the baryon asymmetry of the Universe, three low energy neutrino masses, three physical lepton mixing angles, one Dirac CP phase plus two Majorana CP phases),
the lightest physical neutrino mass is predicted to be zero, while the two Majorana phases are unconstrained (one vanishes),
leaving \textbf{seven observables} currently constrained by data. Thus, there are three degrees of freedom (d.o.f.) in the fit, 
corresponding to the difference between the seven observables and the four input parameters.
Taking into account RG corrections, we estimate $\chi^2 \simeq 1.5-2.6$ for the three d.o.f., depending on the high-energy scale and the type of non-supersymmetric LS model.
We extract allowed ranges of neutrino parameters from our fit data, including the approximate mu-tau
symmetric predictions $\theta_{23}=45^o\pm 1^o$
and $\delta = -90^o \pm 5^o $, which, together with a
normal mass ordering with $m_1=0$, will enable LS models to be tested in future neutrino experiments.
For instance, initial results from NO$\nu$A excluded maximal atmospheric mixing and hence 
were in tension with the model, and future updates from this experiment may provide an early test of the LS.

The layout of the remainder of the paper is as follows.
In \textbf{Section~\ref{LS}} we discuss the LS model.
In \textbf{Section~\ref{RG}} we describe the RG running.
In \textbf{Section~\ref{LG}} we discuss leptogenesis.
In \textbf{Section~\ref{method}} we review our methodology.
In \textbf{Section~\ref{results}} we present our results.
\textbf{Section~\ref{conclusion}} concludes the paper.

\section{The Littlest Seesaw Model}
\label{LS}

The \textbf{seesaw mechanism}~\cite{Minkowski:1977sc, Yanagida:1979ss, Gell-Mann:1979ss, Glashow:1979ss, Mohapatra:1979ia} extends the standard model (SM) with a number of right-handed neutrino singlets $N^{}_{i{\rm R}}$
as, 
\begin{eqnarray}
-{\cal L}^{}_{\rm m} = \overline{L^{}_{\rm L}} \lambda^{}_l H E^{}_{\rm R} + \overline{L^{}_{\rm L}} \lambda^{}_\nu \tilde{H} N^{}_{\rm R} + \frac{1}{2} \overline{N^c_{\rm R}} M^{}_{\rm R} N^{}_{\rm R} + {\rm h.c.} \; ,
\end{eqnarray}
where $L^{}_{\rm L}$ and $\tilde{H} \equiv {\rm i}\sigma^{}_2 H^*$ stand respectively for the left-handed lepton and Higgs doublets, $E^{}_{\rm R}$ and $N^{}_{\rm R}$ are the right-handed charged-lepton and neutrino singlets, $\lambda^{}_l$ and $\lambda^{}_\nu$ are the charged-lepton and Dirac neutrino Yukawa coupling matrices, and $M^{}_{\rm R}$ is the Majorana mass matrix of right-handed neutrino singlets. 
Physical light effective Majorana neutrino masses are generated via the seesaw mechanism, resulting in the light left-handed Majorana neutrino mass matrix
\begin{equation}
 m_\nu = - v^2 \lambda_{\nu} M_R^{-1} \lambda_{\nu}^T\,. 
 \label{eqn:type_1_seesaw}
\end{equation}
The \textbf{2RHN model}
extends the SM by two heavy right-handed neutrino singlets with masses $M_{atm}$ and $M_{sol}$.
In addition, we consider some family symmetry broken by triplet flavons $\phi_i$, whose vacuum alignment will control the structure of the Yukawa couplings. The relevant operators responsible for the Yukawa structure in the neutrino sector are
\begin{equation}
	\frac{1}{\Lambda} \tilde{H} (\overline{L} \cdot \phiatm) N_{\rm atm} + \frac{1}{\Lambda} \tilde{H} (\overline{L} \cdot\phisol )
	N_{\rm sol},
\label{Ynu_flavon}
\end{equation}
where $L$ combines the SU(2) lepton doublets, such that it transforms as a triplet under the family symmetry, while 
$ N_{\rm atm}, N_{\rm sol}$ are the 
right-handed neutrinos $N_R$ and $ H$ is the electroweak scale up-type Higgs SU(2) doublet, the latter two being family symmetry singlets but distinguished by some additional quantum numbers.
The right-handed neutrino Majorana superpotential is typically chosen to give a diagonal mass matrix,
\begin{equation}
	M_{\rm R} = \mathrm{diag}(M_{\rm atm},M_{\rm sol})
	\label{eq:mrr}
\end{equation}
The idea is that \textbf{CSD($n$)} emerges from flavon
vacuum alignments in the effective operators, involving flavon fields $ \phiatm $, $ \phisol $
which are triplets under the flavour symmetry and acquire vacuum expectation values that 
break the family symmetry. 
The subscripts are chosen by noting that $ \phiatm $ correlates with the atmospheric neutrino mass $ m_3 $, and $ \phisol $ with the solar neutrino mass $ m_2 $. CSD($n$) corresponds to the choice of vacuum alignments,
\begin{equation}
	\langle {\phiatm} \rangle= v_{\mathrm{atm}}\pmatr{0 \\ 1 \\ 1}, 
	\qquad \langle {\phisol}\rangle =v_{\mathrm{sol}} \pmatr{1\\n\\(n-2)} \ \ {\rm or} \ \
	 \langle {\phisol}\rangle =v_{\mathrm{sol}} \pmatr{1\\(n-2)\\n}
	\label{CSD(n)}
\end{equation}

where $n$ is a positive integer, and the only phases allowed are in the overall proportionality constants.
Such vacuum alignments are discussed for example in \cite{King:2016yvg}.

In the flavour basis, where the charged leptons and right-handed neutrinos are diagonal,
the {\bf Cases~A, B} are defined by the mass hierarchy $M_{atm}\ll M_{sol}$,
 hence $\widehat{M}^{}_{\rm R} = {\rm Diag}\{ M^{}_{\rm atm}, M^{}_{\rm sol} \}$,
and the structure of the respective Yukawa coupling matrix:
\begin{eqnarray}
{\bf Case~A}:~\lambda^{A}_\nu = \begin{pmatrix}
0 & b e^{{\rm i}\eta/2} \\
a & n b e^{{\rm i}\eta/2} \\
a & (n-2) b e^{{\rm i}\eta/2}
\end{pmatrix} \quad {\rm or} \quad {\bf Case~B}:~\lambda^{B}_\nu = \begin{pmatrix}
0 & b e^{{\rm i}\eta/2} \\
a & (n-2) b e^{{\rm i}\eta/2} \\
a & n b e^{{\rm i}\eta/2}
\end{pmatrix}
\label{eq:Ynu0}
\end{eqnarray}

with $a, b, \eta$ being three real parameters and $n$ an integer. These scenarios were analysed in 
 \cite{King:2016yef,Geib:2017bsw}
with heavy neutrino masses of $M_{\rm atm}^{} = M^{}_1 = 10^{12}~{\rm GeV}$ and $M_{\rm sol}^{} = M_2^{} = 10^{15}~{\rm GeV}$. 

Considering an alternative mass ordering of the two heavy Majorana neutrinos -- $M_{atm}\gg M_{sol}$, and consequently $\widehat{M}^{}_{\rm R} = {\rm Diag}\{ M^{}_{\rm sol}, M^{}_{\rm atm} \}$ -- we have to exchange the two columns of $\lambda^{}_\nu$ in Eq.~(\ref{eq:Ynu0}), namely,
\begin{eqnarray}
{\bf Case~C}: \lambda^{C}_\nu = \begin{pmatrix}
b e^{{\rm i}\eta/2} & 0 \\
n b e^{{\rm i}\eta/2} & a \\
(n-2) b e^{{\rm i}\eta/2} & a
\end{pmatrix} ~~~ {\rm or} ~~~ {\bf Case~D}: \lambda^{D}_\nu = \begin{pmatrix}
b e^{{\rm i}\eta/2} & 0 \\
(n-2) b e^{{\rm i}\eta/2} & a \\
n b e^{{\rm i}\eta/2} & a
\end{pmatrix}\,,
\label{eq:Ynu0I}
\end{eqnarray}

which we refer to as  {\bf Cases~C, D}. For  $M_{\rm atm}^{} = M^{}_2 = 10^{15}~{\rm GeV}$ and $M_{\rm sol}^{} = M_1^{} = 10^{12}~{\rm GeV}$, both these cases were studied in \cite{King:2016yef,Geib:2017bsw}.

Below the right-handed neutrino mass scales, we can 
apply the seesaw formula in Eq.~(\ref{eqn:type_1_seesaw}), for {\bf Cases~A, B, C, D} using the Yukawa coupling matrices $\lambda^{A,B}_{\nu}$ in Eq.~(\ref{eq:Ynu0}) with $M^{A,B}_R=\rm{diag}(M_{atm},M_{sol})$ and  $\lambda^{C,D}_{\nu}$ in Eq.~(\ref{eq:Ynu0I}) with $M^{C,D}_R=\rm{diag}(M_{sol},M_{atm})$, to give (after rephasing) the light neutrino mass matrices in terms of the real parameters $m_a=a^2 v^2/M_{atm}$, $m_b=b^2 v^2/M_{sol}$ with $v=174~\rm{GeV}$:
\begin{equation}
 m^{A,C}_\nu= m_a \begin{pmatrix}
             0 & 0 & 0 \\
             0 & 1 & 1 \\
             0 & 1 & 1
            \end{pmatrix} 
       + m_b \rm{e}^{i \eta} \begin{pmatrix}
                              1 & n & (n-2) \\
                              n & n^2 & n(n-2) \\
                              (n-2) & n(n-2) & (n-2)^2
                             \end{pmatrix}\,,
\label{eq:nu_massA}
\end{equation}
\begin{equation}
 m^{B,D}_\nu= m_a \begin{pmatrix}
             0 & 0 & 0 \\
             0 & 1 & 1 \\
             0 & 1 & 1
            \end{pmatrix} 
       + m_b \rm{e}^{i \eta} \begin{pmatrix}
                              1 &  (n-2) & n \\
                               (n-2) &  (n-2)^2 &  n(n-2)  \\
                              n & n(n-2) & n^2
                             \end{pmatrix}\,.
\label{eq:nu_massB}
\end{equation}

Note the seesaw degeneracy of {\bf Cases~A, C} and {\bf Cases~B, D}, which yield the same effective
neutrino mass matrices, respectively.
Studies which ignore renormalisation group (RG) running effects do not distinguish between these degenerate cases. Of course, in our RG study, the degeneracy is resolved and we have to deal separately with the four physically distinct cases.

The neutrino masses and lepton flavour mixing parameters at the electroweak scale $\Lambda^{}_{\rm EW} \sim \mathcal{O}(1000~{\rm GeV})$ can be derived by diagonalising the effective neutrino mass matrix via
\begin{equation}
 U_{\nu L} m_\nu U^T_{\nu L}=\rm{diag}(m_1,m_2,m_3)\,.
\end{equation}

From a neutrino mass matrix as given in Eqs.~(\ref{eq:nu_massA}) and (\ref{eq:nu_massB}), one immediately obtains normal ordering with $m_1=0$. Furthermore, these scenarios only provide one physical Majorana phase $\sigma$.
As discussed above, we choose to start in a flavour basis, where the right-handed neutrino mass matrix $M_R$ and the charged-lepton mass matrix $M_l$ are diagonal. Consequently, the PMNS matrix is given by $U_{PMNS}=U_{\nu L}^\dagger$. We use the standard PDG parametrisation for the mixing angles, and the CP-violating phase $\delta$. Within our LS scenario, the standard PDG Majorana phase $\varphi_1$ vanishes and $-\varphi_2/2=\sigma$. 

The low-energy phenomenology of \textbf{Case A} has been studied in detail both numerically~\cite{King:2013iva,Bjorkeroth:2014vha} and analytically~\cite{King:2015dvf}, where it has been found that the best fit to experimental data of neutrino oscillations is obtained for $n = 3$ for a particular choice of phase $\eta \approx 2\pi /3$, while for \textbf{Case B} the preferred choice is for $n = 3$ and $\eta \approx -2\pi /3$ \cite{King:2013iva,King:2016yvg}. 
Due to the degeneracy of \textbf{Cases A, C} and \textbf{Cases B, D} at tree level, the preferred choice for $n$ and $\eta$ carries over, respectively.

The prediction for the baryon number asymmetry in our Universe via leptogenesis within \textbf{Case A} has been studied~\cite{Bjorkeroth:2015tsa}, where it was shown that \textbf{Case C} for  positive BAU predicts the CP-violating phase 
to be $\delta\approx 90^o$ which is disfavoured by current global fits to neutrino oscillation data. 
It is straightforward to show that  \textbf{Case B} is disfavoured 
for a similar reason. Therefore, taking into account the positive sign of the 
 BAU, and the present experimentally favoured prediction of $\delta\approx -90^o$, 
one is left with two cases of interest, namely \textbf{Case A} with  $\eta = 2\pi /3$ and \textbf{Case D}
with $\eta = -2\pi /3$, respectively,  where $n = 3$ for both cases. 

These successful cases, which define the two cases of the LS model as discussed in the {\bf Introduction}, 
are summarised below:
\begin{eqnarray}
{\bf Case~A}:~\lambda^{A}_\nu = \begin{pmatrix}
0 & b e^{{\rm i}\pi/3} \\
a & 3 b e^{{\rm i}\pi/3} \\
a &  b e^{{\rm i}\pi/3}
\end{pmatrix} \quad {\rm with} \quad 
M_{\rm R} = \mathrm{diag}(M_{\rm atm},M_{\rm sol})
\label{eq:YnuA}
\end{eqnarray}
\begin{eqnarray}
{\bf Case~D}: \lambda^{D}_\nu = \begin{pmatrix}
b e^{-{\rm i}\pi/3} & 0 \\
 b e^{-{\rm i}\pi/3} & a \\
3 b e^{-{\rm i}\pi/3} & a
\end{pmatrix}
 \quad {\rm with} \quad 
 M_{\rm R} = \mathrm{diag}(M_{\rm sol},M_{\rm atm})
\label{eq:YnuD}
\end{eqnarray}

where in both cases the columns are ordered so that the lighter right-handed neutrino of mass $M_1$
is in the first column and the heavier right-handed neutrino of mass $M_2$ is in the second column,
with $M_1<M_2$. In both cases a normal hierarchy is predicted with $m_1=0$ and 
the physical atmospheric neutrino mass $m_3$ is dominantly controlled by the combination $m_a=a^2 v^2/M_{atm}$, 
while the solar neutrino mass $m_2$ is dominantly controlled by the combination $m_b=b^2 v^2/M_{sol}$,
which is the reason for the notation of the RHN masses used above.
These two cases of the LS model will form the focus of the numerical studies in this paper.

\section{Renormalisation Group Running}
\label{RG}
We suppose that the seesaw theory is defined at some high energy scale such as 
the \textbf{grand unification scale}, or GUT scale, which is denoted by $\Lambda_{GUT}$,
typically around $2.0\times10^{16}$ GeV. In the flavour basis, the charged lepton Yukawa matrix and the Majorana mass matrix $M_R$ are both diagonal. The theory is susceptible to large corrections from \textbf{Renormalisation Group Equations} (RGEs); as such, it is necessary to correctly evolve predictions from the model at $\Lambda_{GUT}$ to a scale at which experimental data is available for testing.

In order to achieve this, the relevant RGEs must be known to some finite loop order; this is only possible in a concrete model such as the Littlest Seesaw, where RGEs follow the SM with modifications in the lepton sector arising from the right-handed neutrinos. As a consequence of approximations made in leptogenesis calculations (see \textbf{Section \ref{LG}}), we always consider very hierarchical Majorana masses - as such, each RH neutrino \emph{must} be integrated out separately as we run our parameters to low scales if we wish to obtain accurate predictions.

We make use of \textbf{effective field theories} (EFTs) below the scale at which the heaviest RH neutrino is integrated out, taking care to ensure matching between each EFT at the appropriate scale:\cite{Antusch:2005gp}
\begin{align}
	M_Z \ll M_1 \ll M_2 \ll \Lambda_{GUT}
	\label{eqn:scales_of_interest}
\end{align}
Here we denote the heaviest RH neutrino mass by $M_2$ and the lightest RH neutrino mass by $M_1$, and discuss calculation of the mass matrix in the three distinct regions relevant to our model. For a given renormalisation scale $\mu$:\\

\underline{$\mu>M_2$}

At this scale, RG effects are due to running of the Yukawa matrix and Majorana mass matrix only (see Eq. \ref{eqn:type_1_seesaw}).\\

\underline{$M_1<\mu<M_2$}

In the intermediate EFT (valid between the scales of the two RH neutrinos) the light neutrino mass matrix will be given by:
\begin{align}
	m_\nu^{(2)} = -v^2\Big[\kappa^{(2)} + \lambda_\nu^{(2)}M_1^{-1}(\lambda_\nu^{(2)})^T\Big]
	\label{eqn:nu_masses_in_intermediary_EFT}
\end{align}
The superscript $(2)$ in Eq. \ref{eqn:nu_masses_in_intermediary_EFT} denotes a matrix that has been altered by integrating out the heaviest RH neutrino. For instance, the matrix $\kappa^{(2)} \propto \lambda_\nu M_2^{-1}\lambda_\nu^{T}$ is the correction for this intermediate EFT, which accounts for the heavier RH neutrino below its mass scale.\\ \\

\underline{$\mu<M_1$}

Below $\mu=M_1$, the theory reduces to the SM with a five-dimensional Weinberg operator modification to the Lagrangian as per Eq. \ref{eqn:lagrangian_for_lowest_eft} :
\begin{align}
	\mathcal{L}_{m}(\mu<M_1) = \frac{1}{2}\kappa^{(1)}\Big(\overline{L^{}_{\rm L}}\tilde{H}\Big)\Big(\tilde{H}^T\overline{L^{c}_{\rm L}}\Big) + h.c.
	\label{eqn:lagrangian_for_lowest_eft}
\end{align}
In this analysis, the one-loop RGEs for the LS are numerically solved from the GUT scale to $M_Z$ using the \texttt{REAP} Mathematica package\cite{Antusch:2005gp}, which ensures correct matching between effective theories and allows us to calculate the light neutrino masses and PMNS mixing parameters in each EFT. More complete discussions of RG running in scenarios such as this are given in \cite{King:2016yef,Geib:2017bsw,Antusch:2005gp}.

\section{Leptogenesis}
\label{LG}
It is a known fact that there is a predominance of matter over antimatter present in the observable Universe, which is thought to have arisen in the very early evolution of our local region. The reason behind this is a question that has been subject to much investigation, but is as yet unanswered. There are various theories that attempt to explain this observed asymmetry today.

The hypothetical out-of-equilibrium process in the expanding Universe through which the number of baryons and antibaryons was effectively fixed is known as \textbf{Big Bang Baryogenesis}\cite{Weinberg:1979,Toussaint:1979}. Traditional SM calculations of this thermal freeze-out of baryons predict equal number densities of baryons and antibaryons in contrast to current observations, which have measured the \textbf{Baryon Asymmetry of the Universe} (BAU) - normalised to entropy density - to be:\cite{Ade:2015xua}
\begin{align}
	Y_B = 0.87\pm0.01 \times10^{-10}
	\label{eqn:BAU/s}
\end{align}
In order for matter and antimatter to be produced at different rates, baryon-generating interactions in the early universe must satisfy the \textbf{Sakharov conditions}\cite{Sakharov:1967} - namely, baryon number violation, charge conjugation (C) and charge-parity (CP) violation, and departure from thermal equilibrium.

The Standard Model allows for \textbf{CP violation} in the weak interactions of quarks and leptons, via the irreducible complex phases in the CKM and PMNS matrices, respectively. This has been studied in the quark sector - specifically, in the neutral kaon and B-meson systems - but is yet to be definitively observed in the neutrino sector.

Nevertheless, there are tantalising hints - this has led to suggestions that, at an even earlier and higher-temperature stage of the Universe, CP was violated in the lepton sector, leading to a lepton-antilepton asymmetry. This is known as \textbf{Leptogenesis}\cite{Fukugita:1986hr}. In this scenario, the lepton asymmetry would then have been communicated to the baryon sector as the Universe evolved via sphaleron processes in the Standard Model that violate conservation of both baryon (B) and lepton (L) number, but conserve the difference B-L.

In addition, the discovery of neutrino oscillations has opened up the possibility of relating the BAU to neutrino properties, thus providing an elegant and comprehensive explanation for the matter-antimatter asymmetry observed today. In these models, leptogenesis relies on extensions of the Standard Model such as different formulations of the seesaw mechanism, and the asymmetry is generated by the decays of the heavy Majorana neutrinos which are the seesaw partners of ordinary neutrinos.

In particular, the simplest version of leptogenesis sees it largely dominated by the interactions and decay of the \textbf{lightest right-handed neutrino}\cite{Buchmuller:2004}. As such, a condition of successful baryogenesis yields constraints on the masses of both light and heavy neutrinos, and provides a means of fixing the mass of the lightest RH neutrino in the Littlest Seesaw model.

A positive $Y_B$ corresponds to a negative \textbf{lepton asymmetry} $Y_{\Delta\alpha}$ in the following way:
\begin{align}
	Y_B=-\frac{12}{37}\sum\limits_{\alpha}Y_{\Delta\alpha}
	\label{eqn:BAU-sphaleron}
\end{align}
where $\alpha$ is a flavour index ($\alpha=e,\mu,\tau$) and $12/37$ is the fraction of B-L asymmetry converted into baryon asymmetry through sphaleron processes\cite{Abada:2006ea}. The \textbf{lepton asymmetry} can be parametrised as\cite{Antusch:2006}:
\begin{align}
	Y_{\Delta\alpha}=\eta_\alpha\epsilon_{1,\alpha}Y^{eq}_{N1}(z\ll1)
	\label{eqn:leptonasym}
\end{align}
where $\eta_\alpha$ is an \textbf{efficiency factor}, $\epsilon_{1,\alpha}$ is the \textbf{decay asymmetry} of the lightest right-handed neutrino into lepton flavour $\alpha$, and $Y^{eq}_{N1}(z\ll1)$ is the \textbf{number density} of the same neutrino at $T\gg M_1$ if it was in thermal equilibrium, normalised to entropy density.

It should be noted that given the previous literature\cite{Antusch:2011nz}, this analysis limited the possible range of masses for the lightest RH neutrino to $10^{9}\leq M_1 \leq 10^{12}$ GeV, in order to enforce the condition of successful leptogenesis in the considered scenario. As such, it is necessary to work in the two-flavour basis, where the tauon Yukawa interactions are in equilibrium, and there will only be two eigenstates - for the tauon flavour and linear combination of muon and electron flavours, respectively.

Thus, Eq. \ref{eqn:leptonasym} will become:
\begin{align}
	Y_{\Delta\alpha}=Y^{eq}_{N1}(z\ll1)\left[\eta_\tau\epsilon_{1,\tau}+\eta_2\left(\epsilon_{1,e}+\epsilon_{1,\mu}\right)\right]
	\label{eqn:leptonasymprime}
\end{align}
where $\eta_2$ is the efficiency factor corresponding to this linear combination of $e$ and $\mu$ flavours, but $\eta_2\neq\eta_e+\eta_\mu$.

In the Boltzmann approximation, the \textbf{number density} in the above expression is given by: \cite{Antusch:2006}
\begin{align}
	Y^{eq}_{N1}(z\ll1)\approx\frac{45}{\pi^4g_*}
	\label{eqn:numberdensity}
\end{align}
and $g_*$ is the number of effective degrees of freedom, which in the Standard Model is $106.75$.

The \textbf{decay asymmetry} into Higgs doublet and left-handed lepton doublet $l_\alpha$ is defined as: \cite{Antusch:2006}
\begin{align}
	\epsilon_{1,\alpha}=\frac{\Gamma_{N_1l_\alpha}-\Gamma_{N_1\bar{l}_\alpha}}{\sum_\alpha(\Gamma_{N_1l_\alpha}+\Gamma_{N_1\bar{l}_\alpha})}
	\label{eqn:decayasym}
\end{align}

where $\Gamma_{N_1l_\alpha}=\Gamma(N_1\rightarrow H+l_\alpha)$ and $\Gamma_{N_1\bar{l}_\alpha}=\Gamma(N_1\rightarrow H^*+\bar{l}_\alpha)$ are the decay rates into particles and antiparticles, respectively. The decay asymmetry is $0$ at tree level but arises at 1-loop level, becoming: \cite{Covi:1996}
\begin{align}
	\epsilon_{1,\alpha}=\frac{1}{8\pi}\frac{\sum_{J=2,3}\textrm{Im}[(\lambda^\dag_\nu)_{1\alpha}[\lambda^\dag_\nu\lambda_\nu]_{1J}(\lambda^T_\nu)_{J\alpha}]}{(\lambda^\dag_\nu\lambda_\nu)_{11}}\times g^{SM}\left(\frac{M^2_J}{M^2_1}\right)
	\label{eqn:decayasym2}
\end{align}

where $\lambda_\nu$ denotes the Yukawa matrix and the loop function $g$ in the SM is given by: \cite{Antusch:2006}
\begin{align}
	g^{SM}(x)=\sqrt{x}\left[\frac{1}{1-x}+1-(1+x)\ln\frac{1+x}{x}\right]
	\label{eqn:loopg}
\end{align}
In the case of Eq. \ref{eqn:decayasym2}, $x=\cfrac{M^2_J}{M^2_1}$. In addition, for the Littlest Seesaw, $J_{2,3}$ can be simplified to $J_2$, as only two RH neutrinos are present in the model. Consequently, the sums will disappear from the equation. It should be noted the factor $J_2$ will always appear due to the fact that $N_2$ is in the loop for the decay of $N_1$.\cite{Covi:1996}

As such, there will be some sensitivity in our analysis to the mass of the heavier RH neutrino. However, it can be easily proven that the dominant contribution will come from the lighter neutrino mass: under the assumption of a hierarchical limit $M_1\ll M_2$, Eq. \ref{eqn:loopg} can be approximated as $g^{SM}(x)\approx-\cfrac{3}{2\sqrt{x}}$ \footnote{It should be noted that in our analysis we do use the full expression for $g^{SM}$ given in Eq. \ref{eqn:loopg} and also include the RG running effects on leptogenesis down to $M_2$. The approximations discussed here are purely for the purposes of illustrating the respective contributions of $M_1$ and $M_2$.}. Then, for the \textbf{Case A} Yukawa matrix given in Eq. \ref{eq:Ynu0},  the individual flavour-dependent asymmetries become at $ \Lambda_{GUT}$:
\begin{align}
	\begin{aligned}
		\epsilon^A_{1,e}&=0\\
		\epsilon^A_{1,\mu}&\approx -\frac{3}{16\pi}\frac{M_1}{M_2}n(n-1)b^2\sin{\eta} \label{epsilonA}\\
		\epsilon^A_{1,\tau}&\approx -\frac{3}{16\pi}\frac{M_1}{M_2}(n-1)(n-2)b^2\sin{\eta}
	\end{aligned}
\end{align}

As stated in \textbf{Section \ref{LS}}, the quantity $b^2/M_2$ is proportional to $m_b$ - in our analysis, this is kept to within an order of magnitude for \textbf{Case A} to obtain favourable neutrino observables at low energies. Thus, the mass of the lighter RH neutrino $M_1$ will have a larger effect in the above decay asymmetries, and this is therefore the parameter that will be most sensitive to the BAU.

Analogously, the decay asymmetries in \textbf{Case D} would reduce to the following:
\begin{align}
	\begin{aligned}
		\epsilon^D_{1,e}&=0\\
		\epsilon^D_{1,\mu}&\approx \frac{3}{8\pi}\frac{M_1}{M_2}\frac{(n-1)(n-2)}{(2n^2-4n+5)}a^2\sin{\eta}
		 \label{epsilonD} \\
		\epsilon^D_{1,\tau}&\approx \frac{3}{8\pi}\frac{M_1}{M_2}\frac{n(n-1)}{(2n^2-4n+5)}a^2\sin{\eta}
	\end{aligned}
\end{align}

In this case, it is the quantity $a^2/M_2$ which is kept approximately fixed; hence $M_1$ will still be the most sensitive parameter with respect to the decay asymmetry.

When there is a vanishing initial abundance ($N^{in}_{N_1}=0$) the \textbf{efficiency factor} will be composed of a negative and positive contribution:,\cite{Buchmuller:2004,Blanchet:2007}
\begin{align}
	\eta_{\alpha}=\eta_-(K_\alpha,P_{1\alpha}^0)+\eta_+(K_\alpha,P_{1\alpha}^0) 
	\label{eqn:2cont}
\end{align}
where $K_\alpha$ are \textbf{decay parameters} and $P^0_{1\alpha}$ are the \textbf{tree level branching ratios}.

As before, when working in the two-flavour basis, the above expression will need to be treated separately for each of the two eigenstates:
\begin{align}
	\begin{aligned}
		\eta_{\tau}&=\eta_-(K_\tau,P_{1\tau}^0)+\eta_+(K_\tau,P_{1\tau}^0)\\
		\eta_2&=\eta_-(K_2,P_{12}^0)+\eta_+(K_2,P_{12}^0)
	\end{aligned}
	\label{eqn:2cont_2}
\end{align}
The \textbf{negative contribution} comes from an initial stage where $N_{N_1}\leq N_{N_1}^{eq}$ and $z\leq z_{eq}$. It is approximated by: \cite{Blanchet:2007}
\begin{align}
	\eta_-(K_\alpha,P_{1\alpha}^0)\simeq-\frac{2}{P^0_{1\alpha}}e^{-\cfrac{3\pi K_{\alpha}}{8}}\left(e^{\cfrac{P^0_{1\alpha}}{2}N_{N_1}(z_{eq})}-1\right)
	\label{eqn:negcont}
\end{align}
where, in the above expression:

\begin{itemize}
\item The \textbf{$N_1$ abundance} at $z_{eq}$ is defined as: \cite{Blanchet:2007}
\begin{align}
	N_{N_1}(z_{eq})\simeq\overline{N}(K_\alpha)\equiv\cfrac{N(K_\alpha)}{\left(1+\sqrt{N(K_\alpha})\right)^2}
	\label{eqn:N1abundance}
\end{align}
and $N(K_\alpha)=3\pi\cfrac{K_\alpha}{4}$, where $K_\alpha$ is the \textbf{decay parameter}\cite{Dev:2017trv}. For the flavour-dependent case this will become $K_2$, combining the $e$ and $\mu$ decay parameters. For each flavour, the individual decay parameter will be given by:
\begin{align}
	K_{\alpha}=(\lambda_\nu^\dag)_{1\alpha}(\lambda_\nu)_{\alpha 1}\frac{v^2}{m^*_{SM}M_1}
	\label{eqn:decayparam}
\end{align}
where $m^*_{SM}\approx1.08*10^{-3}$eV is the equilibrium neutrino mass \cite{Buchmuller:2004} and $v$ is the SM Higgs vacuum expectation value ($v=246/\sqrt{2}$). Thus, $K_2$ will simply be the sum of the contributions from $\alpha=e,\mu$, such that $K_2=\sum\limits_\alpha^{e,\mu}K_\alpha$.

\item The \textbf{tree level branching ratios} are given by:
\begin{align}
	P^0_{1\alpha}\simeq\frac{|\lambda_{\alpha 1}|^2}{(\lambda^\dag\lambda)_{11}}
	\label{eqn:branchratios}
\end{align}
and similarly, $P_{12}^0$ in Eq. \ref{eqn:2cont_2} refers to $P_{12}^0=\sum\limits_\alpha^{e,\mu}P_{1\alpha}^0$ in our two-flavour basis.
\end{itemize}
The \textbf{positive contribution} will correspond to the stage when $N_{N_1}\geq N_{N_1}^{eq}$ and $z\geq z_{eq}$. It can be approximated by: \cite{Blanchet:2007}
\begin{align}
	\eta_+(K_\alpha,P_{1\alpha}^0)\simeq\frac{2}{z_B(K_{\alpha})K_{\alpha}}\left(1-e^{-\cfrac{K_{\alpha}z_B(K_{\alpha})N_{N_1}(z_{eq})}{2}}\right)
	\label{eqn:poscont}
\end{align}
where the same definitions hold as for the negative contribution and in addition the \textbf{freeze-out parameter} is given by $z_B(K_{\alpha})\simeq2+4K_{\alpha}^{0.13}e^{-\frac{2.5}{K_{\alpha}}}$. \cite{Blanchet:2006}

In this way, all of the necessary terms can be determined to calculate a prediction of the BAU at each parameter point scanned over in the model. The results of this will be shown in \textbf{Section \ref{results}}.

The inclusion of leptogenesis in this analysis immediately invalidates \textbf{Cases B} and \textbf{C} in the Littlest Seesaw. A general seesaw mechanism in CSD($n$) assuming two right-handed neutrinos will involve a single phase $\eta$, which provides the link between the neutrino oscillation phase $\delta_{\text{CP}}$ and the CP violating phenomenon responsible for generating a matter-antimatter asymmetry in leptogenesis models \cite{King:2002}. \footnote{This phase is not to be confused with the efficiency factor $\eta$ included in the above calculations.}

Following the derivation set out in \cite{Antusch:2006}, this relation can be seen as $Y_B\propto\pm\sin{\eta}$, where the positive sign will correspond to the scenario in which $M_{\text{atm}}\ll M_{\text{sol}}$ (in our model, \textbf{Cases A} and \textbf{B}) and the negative to the scenario $M_{\text{sol}}\ll M_{\text{atm}}$ (\textbf{Cases C} and \textbf{D}). As the resulting BAU observed is required to be 
positive, it is clear that $\eta$ must therefore be positive in the first category and negative in the second.

However, as discussed in \textbf{Section \ref{LS}}, the phenomenology of each of the LS cases has been studied extensively, and \textbf{Case B} has been shown to prefer a value of $\eta\approx-2\pi/3$, corresponding to 
the current experimental hint for $\delta_{CP}\sim-\pi/2$,
while \textbf{Case C} prefers $\eta\approx2\pi/3$ for similar reasons.
Given the required positive BAU, we see therefore that \textbf{Cases B} and \textbf{C} are in conflict with current hints for 
$\delta_{CP}\sim-\pi/2$.
Thus, when leptogenesis calculations are included in the analysis, 
we consider only \textbf{Cases A} and \textbf{D} 
- given in Eqs.~\ref{eq:YnuA} and \ref{eq:YnuD}, respectively - which can be seen from Eqs.~\ref{epsilonA},~\ref{epsilonD} to lead to 
negative lepton asymmetries and hence a positive final baryon asymmetry.

\section{Methodology}
\label{method}
As discussed above, we focus exclusively on 
\textbf{Cases A} and \textbf{D} 
in Eqs.~\ref{eq:YnuA} and \ref{eq:YnuD}.
Each of these cases involve just four real free parameters at high energies which will predict the entire neutrino sector and the BAU from leptogenesis - two \textbf{Yukawa parameters} $a$ and $b$ and two \textbf{RH neutrino masses} $M_{atm}$ and $M_{sol}$. In this work, we determine the RH neutrino masses and Yukawa parameters through a fit to low-scale experimental data, using the $\bm{\chi^2}$ \textbf{function} as a goodness-of-fit measure:
\begin{align}
	\chi^2 = \sum_{i=1}^N\Bigg( \frac{P_i(x)-\mu_i}{\sigma^i} \Bigg)^2
	\label{eqn:chi2_definition}
\end{align}

\textbf{Low-energy predictions} of the model denoted $P_i(x)$ are fully determined by our set of four parameters, collectively labelled $x$.
\begin{spreadlines}{8pt}
\begin{gather}
		x = (a,b,M_{atm},M_{sol})\\
		\mu_i = (\sin^2\theta_{12},\sin^2\theta_{13},\sin^2\theta_{23},\Delta m_{12}^2,\Delta m_{13}^2,\delta,Y_B)
	\label{eqn:parameter_set}
\end{gather}
\end{spreadlines}

The LS is tested against the recent \textbf{global fit values} of neutrino data from NuFit3.2 \cite{Esteban:2016qun} and the measured value of the \textbf{BAU} from the Planck satellite's 2015 data\cite{Ade:2015xua}. $\sigma_i$ in Eq. \ref{eqn:chi2_definition} corresponds to the 1$\sigma$ bounds for each of the experimentally measured values $\mu_i$. If the data follows a Gaussian distribution, these bounds are simply the standard deviations with respect to the central values.
\bigskip


\begin{table}[H]
\centering
\begin{tabular}{lrr}
	\toprule \addlinespace
		\textbf{Observable} & \textbf{NuFit3.2($\boldsymbol{\pm}$1$\boldsymbol{\sigma}$)} & \textbf{Assumed Values($\boldsymbol{\pm}$1$\boldsymbol{\sigma}$)} \\ \addlinespace
	\midrule \addlinespace
		$\theta_{12}/^\circ$ & $33.62_{-0.76}^{+0.78}$ & $33.62_{-0.76}^{+0.76}$ \\ \addlinespace
		$\theta_{13}/^\circ$ & $8.54_{-0.15}^{+0.15}$ & $8.54_{-0.15}^{+0.15}$ \\ \addlinespace
		$\theta_{23}/^\circ$ & $47.2_{-3.6}^{+1.9}$ & $46.4_{-2.8}^{+2.8}$ \\ \addlinespace
		$\delta/^\circ$ & ${-126}^{+43}_{-31}$ & ${-120}^{+37}_{-37}$ \\ \addlinespace
		$\Delta m_{12}^2/\text{eV}^2$ & $7.40_{-0.20}^{+0.21} \times 10^{-5}$ & $7.40_{-0.20}^{+0.20} \times 10^{-5}$ \\ \addlinespace
		$\Delta m_{13}^2/\text{eV}^2$ & $2.494_{-0.031}^{+0.033} \times 10^{-3}$ & $2.494_{-0.031}^{+0.031} \times 10^{-3}$ \\ \addlinespace
	\midrule \addlinespace
	$Y_B$ & $0.87_{-0.01}^{+0.01} \times 10^{-10}$ & $0.87_{-0.01}^{+0.01} \times 10^{-10}$ \\ \addlinespace
	\bottomrule
\end{tabular}
\caption{Global fit values from NuFit3.2\cite{Esteban:2016qun} in the case of normal ordering, along with latest 1$\sigma$ bounds on $Y_B$ from Planck satellite\cite{Ade:2015xua}. The third column details the assumed values which we actually use in our analysis, which respect the one sigma asymmetric ranges in NuFit3.2.}
\label{tab:experimental}
\end{table}
We also calculate the \textbf{reduced} $\bm{\chi^2}$ for each parameter point:
\begin{align}
\chi_\nu^2 = \frac{\chi^2}{\dof}
\label{eqn:reduced_chi2_def.}
\end{align}

where $\dof$ is the number of degrees of freedom - the number of observables minus the number of parameters in our analysis ($\equiv$3 here). A model is said to be a good fit to data if $\chi^2_\nu \simeq 1$ and any model that deviates significantly from this is said to be a poor fit. In \textbf{Table~\ref{tab:experimental}} it can be seen that the distributions for some experimentally measured observables are asymmetric; hence, to keep the analysis clean and simple, we approximate the $\chi^2$ for each parameter point in the following manner:

To be conservative, we approximate that the observables conform to a symmetric Gaussian distribution, where the central value are those given in the table and we use the smaller of the quoted uncertainties in our calculations for all constraints except $\theta_{23}$ and $\delta$. For these two observables, we approximate the distribution as Gaussian with its central value located in the middle of the quoted $1\sigma$ range from \textbf{Table \ref{tab:experimental}}. Thus, the modifications from the true experimental values stated in the table are as follows: $\theta_{23}={46.35^\circ}_{-2.75^\circ}^{+2.75^\circ}$, $\delta={120^\circ}_{-37^\circ}^{+37^\circ}$. 

In this way, the $1\sigma$ range is preserved and we do not vastly overestimate the $\chi^2$ contributions from these observables relative to the rest. However, it is important to note that this method will always underestimate the $\chi^2$ slightly.

\begin{figure}[H]
	\centering
	\begin{tikzpicture}[node distance = 1.8cm, auto]
	\node [cloud1] (init) {Parameters};
	\node [block,below of=init] (mats) {Matrices};
	\node [block,below of=mats] (REAP) {RG Running: \texttt{REAP}\cite{Antusch:2005gp}};
	\node [block0,below of=REAP,left of=REAP] (obs) {Predictions};
	\node [block0,below of=REAP,right of=REAP] (calc) {$\chi^2$};
	\node [io, above of=init] (scan) {Overarching Grid Scan};

	\path [line] (init) -- (mats);
	\path [line] (mats) -- (REAP);
	\path [line] (REAP) -| (obs);
	\path [line] (obs) -- (calc);
	\draw [dashed, black] (scan) -- (init);
	\end{tikzpicture}
	\caption{Parameter scan and data flow for a single point. $\chi^2$ denotes calculation of the goodness-of-fit test against experimental data.}
	\label{fig:flowchart}
\end{figure}
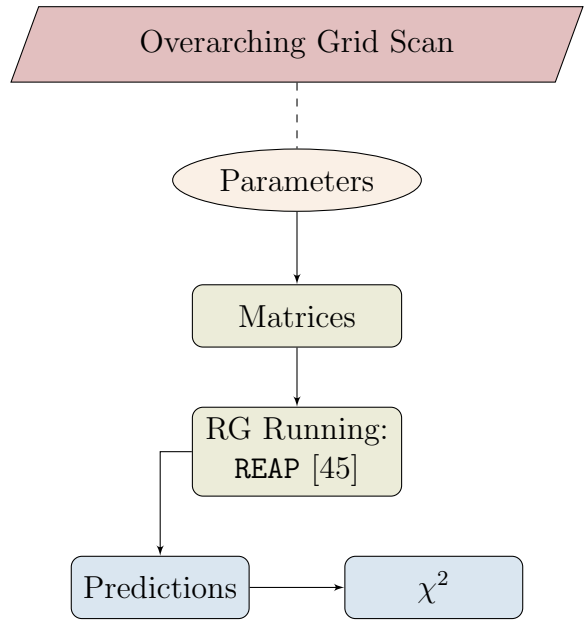

The general method used to test each parameter point is outlined in \textbf{Figure \ref{fig:flowchart}}. We scan over the four input parameters with a predefined step for each one, with ranges initially informed by  previous analysis of the LS\cite{Geib:2017bsw}. A \textbf{four-dimensional grid} is created, and at each point of this grid - corresponding to a particular combination of parameters - the values for each parameter are inserted into the relevant matrices to define a single parameter point at the GUT scale.\\

The Yukawa matrices and Majorana mass matrix are run down from the GUT scale to $M_Z$, using the Mathematica program \texttt{REAP}\cite{Antusch:2005gp}, to integrate out the RH neutrinos and ensure that the relevant EFTs are correctly matched at appropriate scales (see \textbf{Section \ref{RG}}). We extract neutrino and leptogenesis predictions, then calculate the $\chi^2$ using the method discussed above. At each stage of our grid scan we reduce the tested bounds of every input parameter, iteratively scanning more finely in all parameters to find the best fit point with a stable global minimum $\chi^2$.

It is worth noting here that the conventions used by \texttt{REAP} differ from those given here and in other literature. We have defined our Yukawa matrices in \textbf{Section \ref{LS}} in so-called left-right (LR) convention. This means that the matrix is defined with the SU(2) lepton doublet on the left-hand side of the term in the Lagrangian. Conversely, \texttt{REAP} takes only matrices defined in RL convention. Standard PDG parametrisation is used for the mixing angles with the exception of the CP-violating phase $\delta$. NuFit denotes that $\delta\in[-\pi,\pi]$, whereas \texttt{REAP} uses $\delta_{\texttt{REAP}}\in[0,2\pi]$. The conversions can be summarised as the following:
\begin{gather}
(\lambda_{\nu})_{\texttt{REAP}} = \lambda_\nu^\dagger\quad ,\quad\delta_{\texttt{REAP}} = \delta + \pi
\end{gather}

\section{Results}
\label{results}
The results in this section are for 
\textbf{Cases A} and \textbf{D} 
in Eqs.~\ref{eq:YnuA} and \ref{eq:YnuD}.
Each of these cases involve just four real free parameters at high energies which will predict the entire neutrino sector and the BAU from leptogenesis - two real and positive \textbf{Yukawa parameters} $a$ and $b$ and two 
real and positive \textbf{RH neutrino masses} $M_{atm}$ and $M_{sol}$.
Following the method outlined in \textbf{Section \ref{method}}, the main results are summarised in \textbf{Table \ref{tab:method_2_results}}, where we outline \textbf{best fit points} for each of the two cases for two different GUT scales.
\bigskip


\begin{table}[H]
	\centering
	\begin{tabular}{lrrrr}
	\toprule \addlinespace
	& \textbf{Case A1} & \textbf{Case D1} & \textbf{Case A2} & \textbf{Case D2} \\ \addlinespace
	\midrule \addlinespace
	$\Lambda_{GUT}/\text{GeV}$ & $1.0\times10^{16}$ & $1.0\times10^{16}$ & $2.0\times10^{16}$ & $2.0\times10^{16}$ \\ \addlinespace
	\midrule \addlinespace
	$M_{atm}/\text{GeV}$ & $5.10\times10^{10}$ & $1.59\times10^{12}$ & $5.05\times10^{10}$ & $1.36\times10^{13}$ \\ \addlinespace
	$M_{sol}/\text{GeV}$ & $3.28\times10^{14}$ & $1.08\times10^{10}$ & $5.07\times10^{13}$ & $1.06\times10^{10}$ \\ \addlinespace
	$a$ & $0.00817$ & $0.0456$ & $0.00806$ & $0.135$ \\ \addlinespace
	$b$ & $0.215$ & $0.00117$ & $0.0830$ & $0.00116$ \\ \addlinespace
	\midrule \addlinespace
	$\chi^2$/d.o.f. & $1.51/3$ & $2.64/3$ & $1.75/3$ & $2.07/3$ \\ \addlinespace
	\bottomrule
	\end{tabular}
	\caption{Benchmark Point Parameters}
	\label{tab:method_2_results}
\end{table}

\textbf{Case A1} (\textbf{A2} )refers to the benchmark point obtained from a full fit of \textbf{Case A} with the GUT scale fixed at $1.0\times10^{16}$ GeV ($2.0\times10^{16}$ GeV), and analogously for the \textbf{Case D}
best fit points. From these benchmark points we can conclude that both cases of the LS give excellent fits to the experimental data, and also that the $\chi^2$/d.o.f. for each benchmark point are too close to degeneracy to say that any one is preferred over the others, only that they all give a good fit to data.

Here we remind the reader that due to the approximations in our method  (especially our symmetric treatment of the asymmetric errors, of particular relevance for the atmospheric angle)
the $\chi^2$ is always somewhat underestimated. However, although the treatment of asymmetric errors is beyond the scope
of our analysis, our method does respect the asymmetric one sigma ranges of all the observables
(see our \textbf{Assumed Values} in \textbf{Table~\ref{tab:experimental}}), and so any uncertainties in our method should be within one sigma accuracy.

\subsection{Observables and $\chi^2$ Contributions}

In order to explore these results in more detail we take the benchmark points from \textbf{Table \ref{tab:method_2_results}} and look more closely at the predictions for \textbf{experimental observables} made by each of them.
\bigskip

\begin{table}[H]
	\centering
	\begin{tabular}{lrrrrr}
	\toprule \addlinespace
	& \textbf{Case A1} & \textbf{Case D1} & \textbf{Case A2} & \textbf{Case D2} &  \emph{\textbf{Experiment}}\cite{Esteban:2016qun,Ade:2015xua}\\ \addlinespace
	\midrule \addlinespace
	$\theta_{12}/^{\circ}$ & $34.20$ & $34.32$ & $34.30$ & $34.33$ & $33.62_{-0.76}^{+0.78}$\\ \addlinespace
	$\theta_{13}/^{\circ}$ & $8.58$ & $8.64$ & $8.59$ & $8.59$ & $8.54_{-0.15}^{+0.15}$\\ \addlinespace
	$\theta_{23}/^{\circ}$ & $45.33$ & $44.24$ & $45.62$ & $44.24$ & $47.2_{-3.6}^{+1.9}$\\ \addlinespace
	$\Delta {m_{21}}^2/10^{-5}\text{eV}^2$ & $7.43$ & $7.33$ & $7.36$ & $7.34$ &  $7.40_{-0.20}^{+0.21}$\\ \addlinespace
	$\Delta {m_{31}}^2/10^{-3}\text{eV}^2$ & $2.49$ & $2.48$ & $2.50$ & $2.50$ &  $2.494_{-0.031}^{+0.033}$\\ \addlinespace
	$\delta/^{\circ}$ & $-89.0$ & $-93.2$ & $-87.4$ & $-93.0$ & ${-126}^{+43}_{-31}$\\ \addlinespace
	$Y_B/10^{-10}$ & $0.860$ & $0.860$ & $0.860$ & $0.861$ & $0.87_{-0.01}^{+0.01}$\\ \addlinespace
	\midrule \addlinespace
	$\chi^2$/d.o.f. & $1.51/3$ & $2.64/3$ & $1.76/3$ & $2.07/3$ & -\\ \addlinespace
	\bottomrule
	\end{tabular}
	\caption{Benchmark Point Observables}\label{tab:observables}
\end{table}
	
The predictions for $\theta_{23}$ and $\delta$ are particularly interesting. We see that regardless of the Case studied, the Littlest Seesaw does indeed yield close to maximal atmospheric mixing. Likewise, the CP-violating phase is consistently predicted to be in the vicinity of $-\pi/2$. 

\begin{figure}[H]
\centering
\begin{subfigure}[b]{0.495\linewidth}
	\includegraphics[width=\linewidth]{{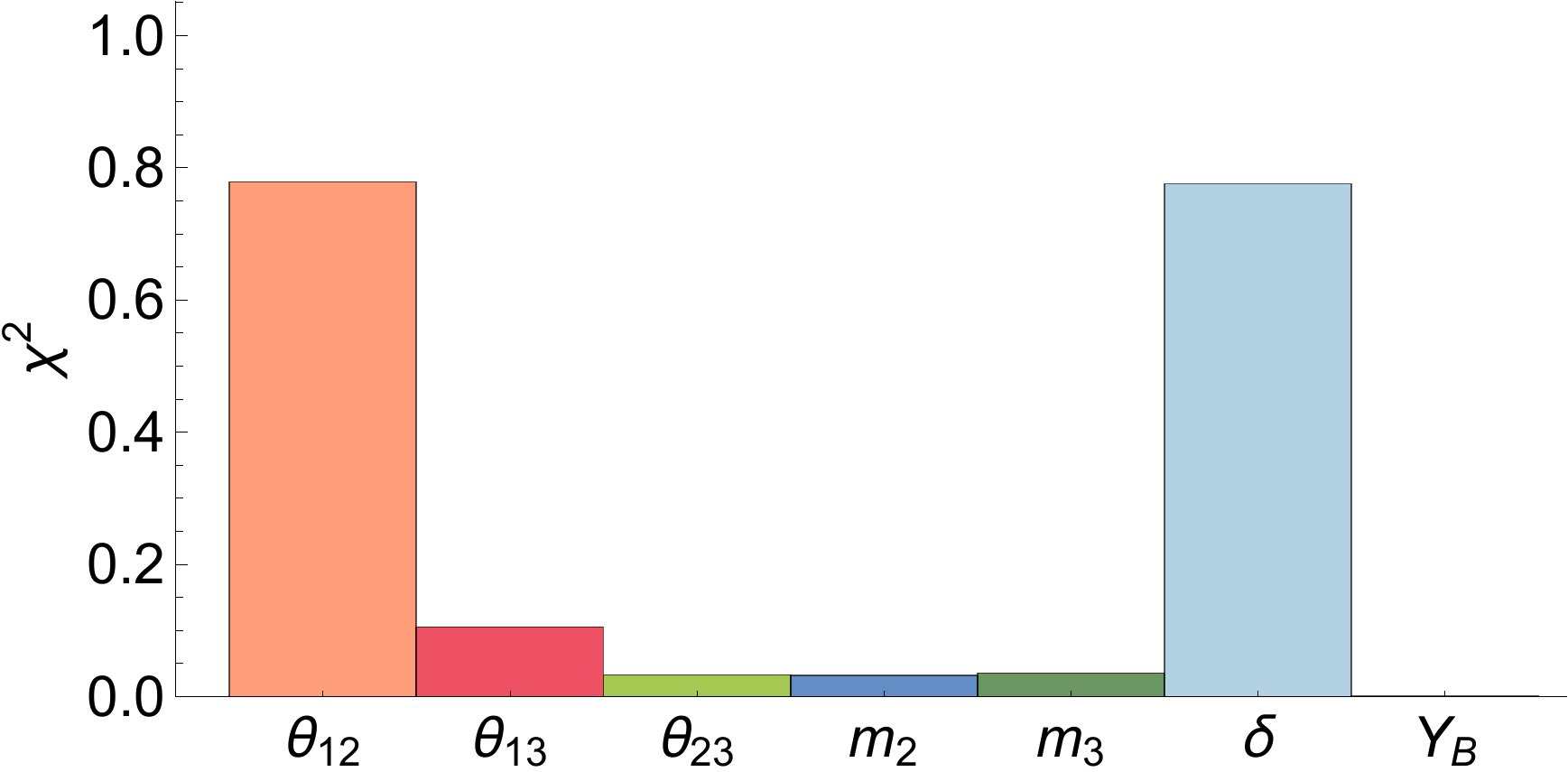}}
	\caption{\textbf{Case A2}}
	\label{fig:A_pulls}
\end{subfigure}
\begin{subfigure}[b]{0.495\linewidth}
	\includegraphics[width=\linewidth]{{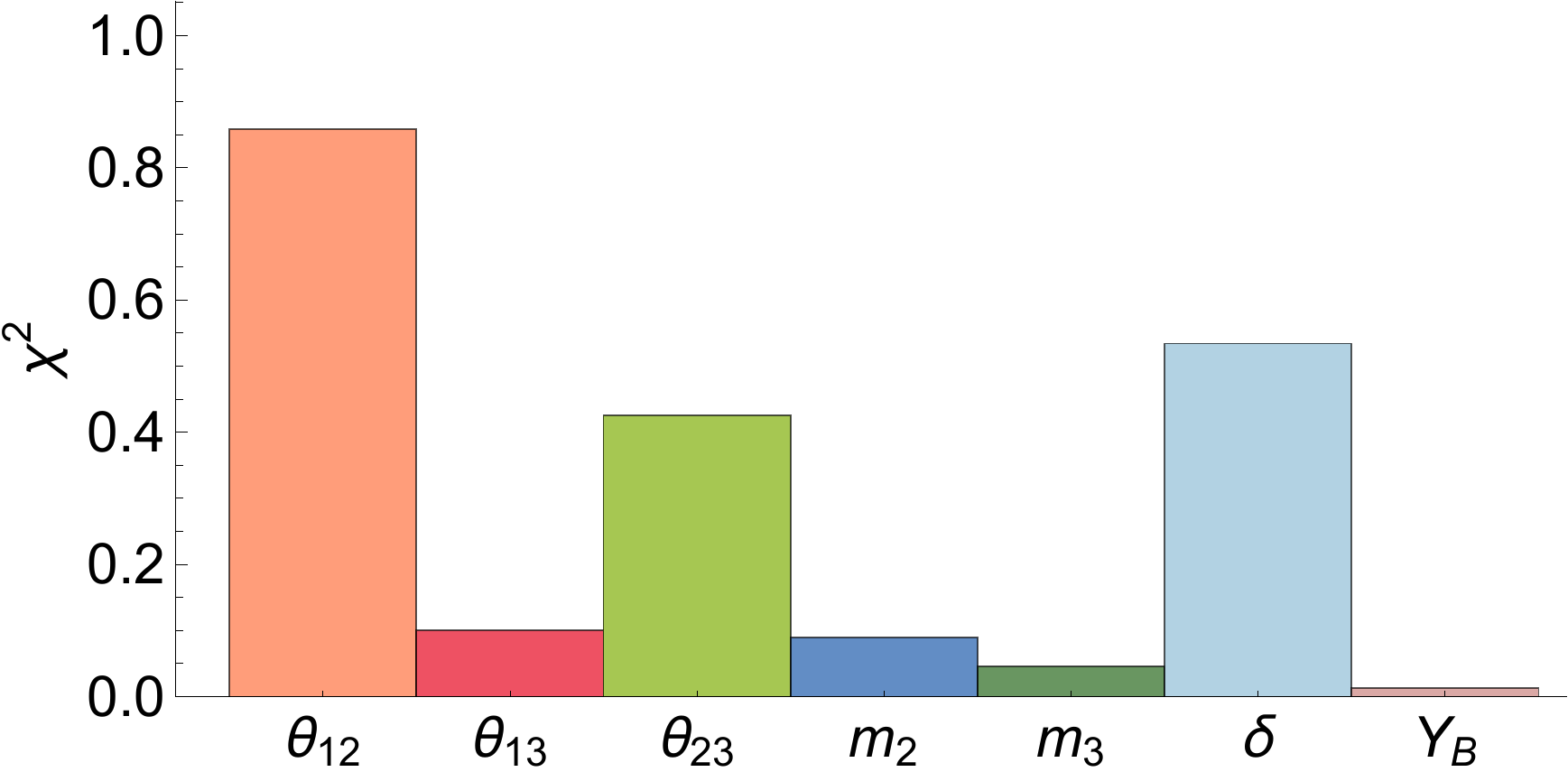}}
	\caption{\textbf{Case D2}}
	\label{fig:D_pulls}
\end{subfigure}
\caption{$\chi^2$ contribution of each observable for \textbf{Case A2} and \textbf{D2} best fit points.}
\label{fig:pulls}
\end{figure}

In order to be concise, we concentrate on \textbf{Case A2} and \textbf{Case D2} for the remainder of the results shown. \textbf{Figure \ref{fig:pulls}} depicts the \textbf{contributions} to the total $\chi^2$ of each observable.

$\theta_{12}$ can be seen to exert a large pull over the data for both cases. This observable is always fixed close to $33^\circ-34^\circ$, resulting from a sum rule that can be derived from the model itself:
\begin{gather}
	\tan\theta_{12} = \frac{1}{\sqrt{2}}\sqrt{1-3\sin^2\theta_{13}}
\end{gather}
A more complete discussion of analytic predictions of the LS model is given in~\cite{King:2015dvf,King:2016yvg}.

\subsection{RG Effects in Benchmark Points}
\begin{figure}[H]
	\centering
		\includegraphics[width=0.49\textwidth]{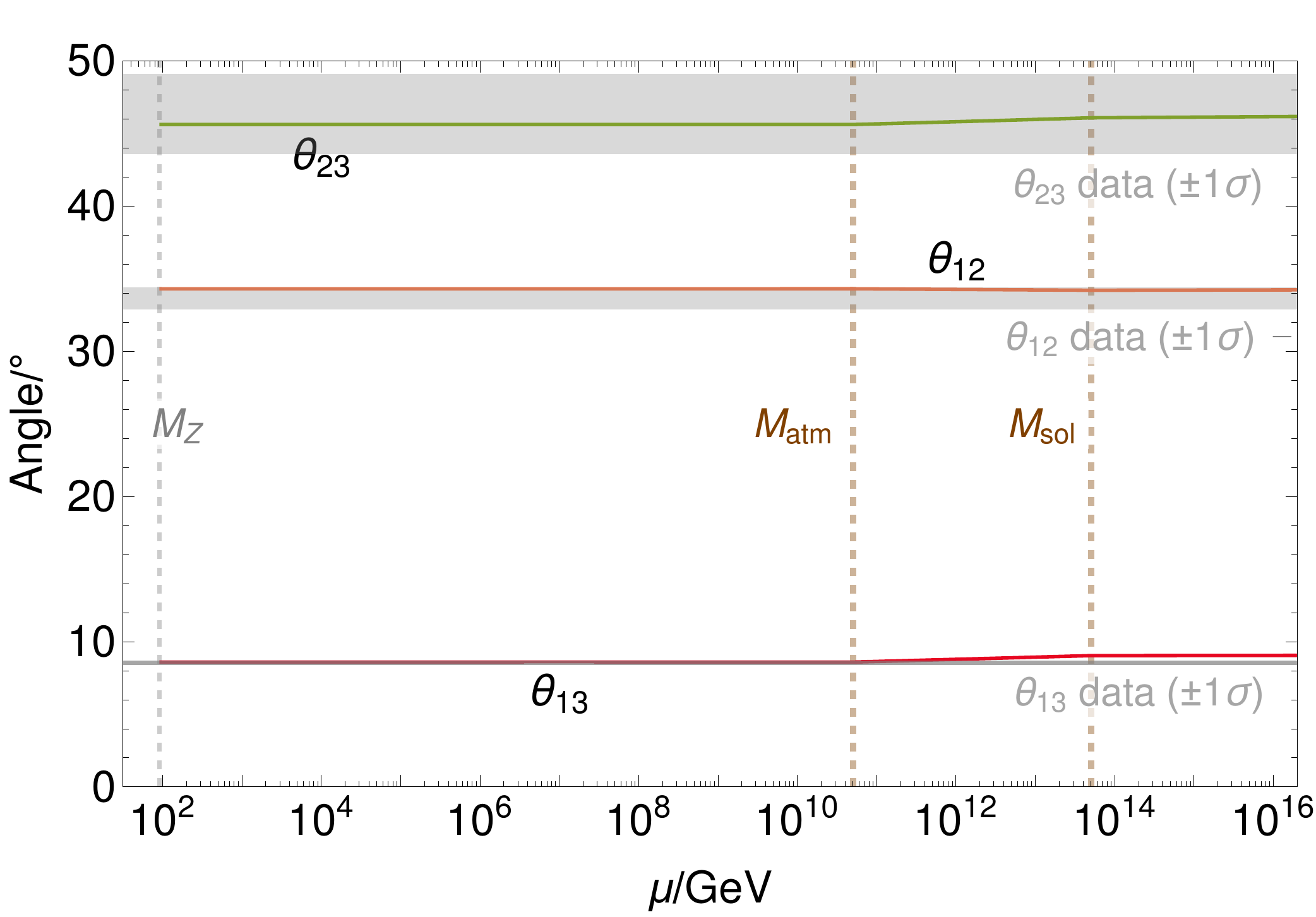}
	\includegraphics[width=0.49\textwidth]{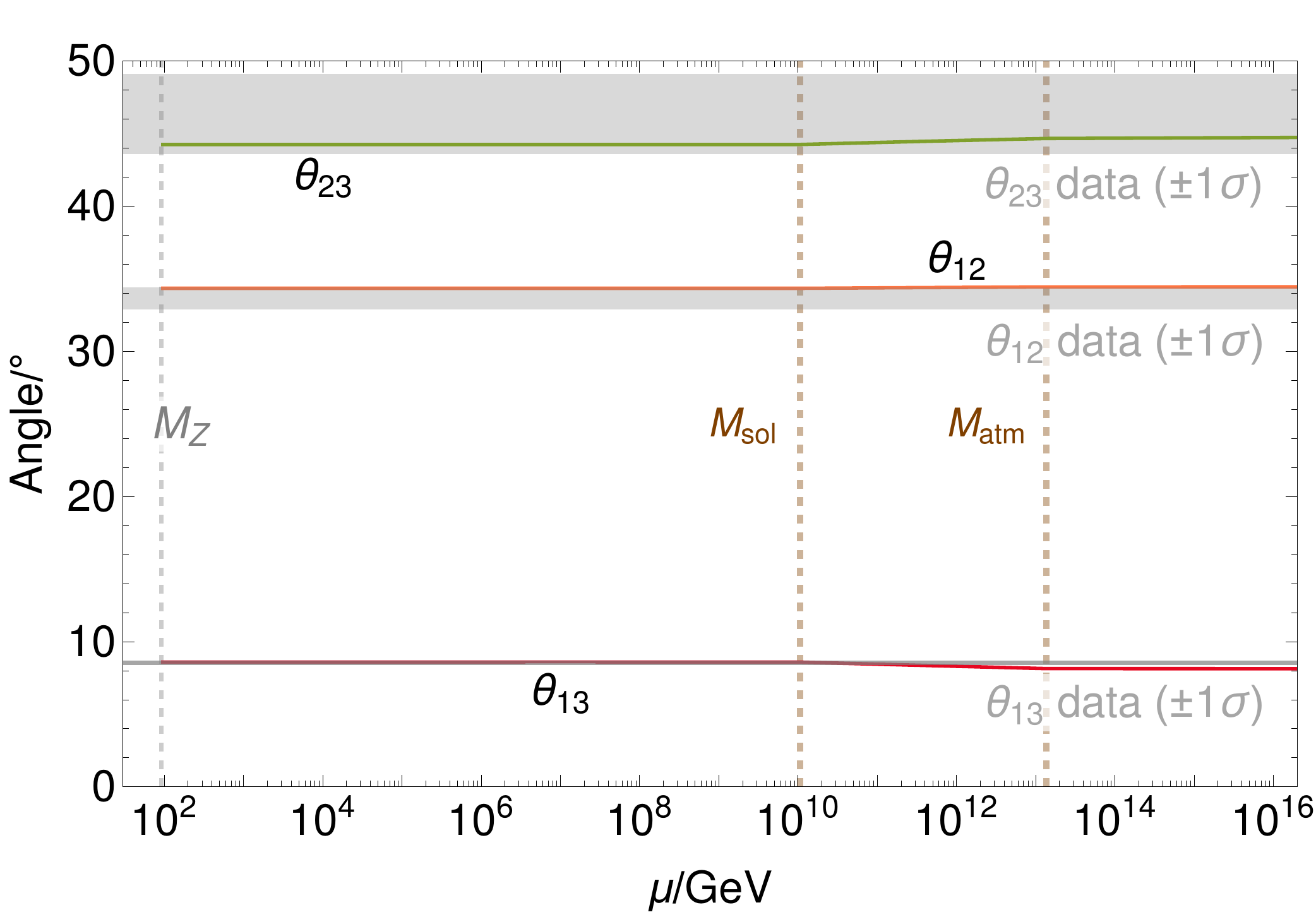}
	\includegraphics[width=0.49\textwidth]{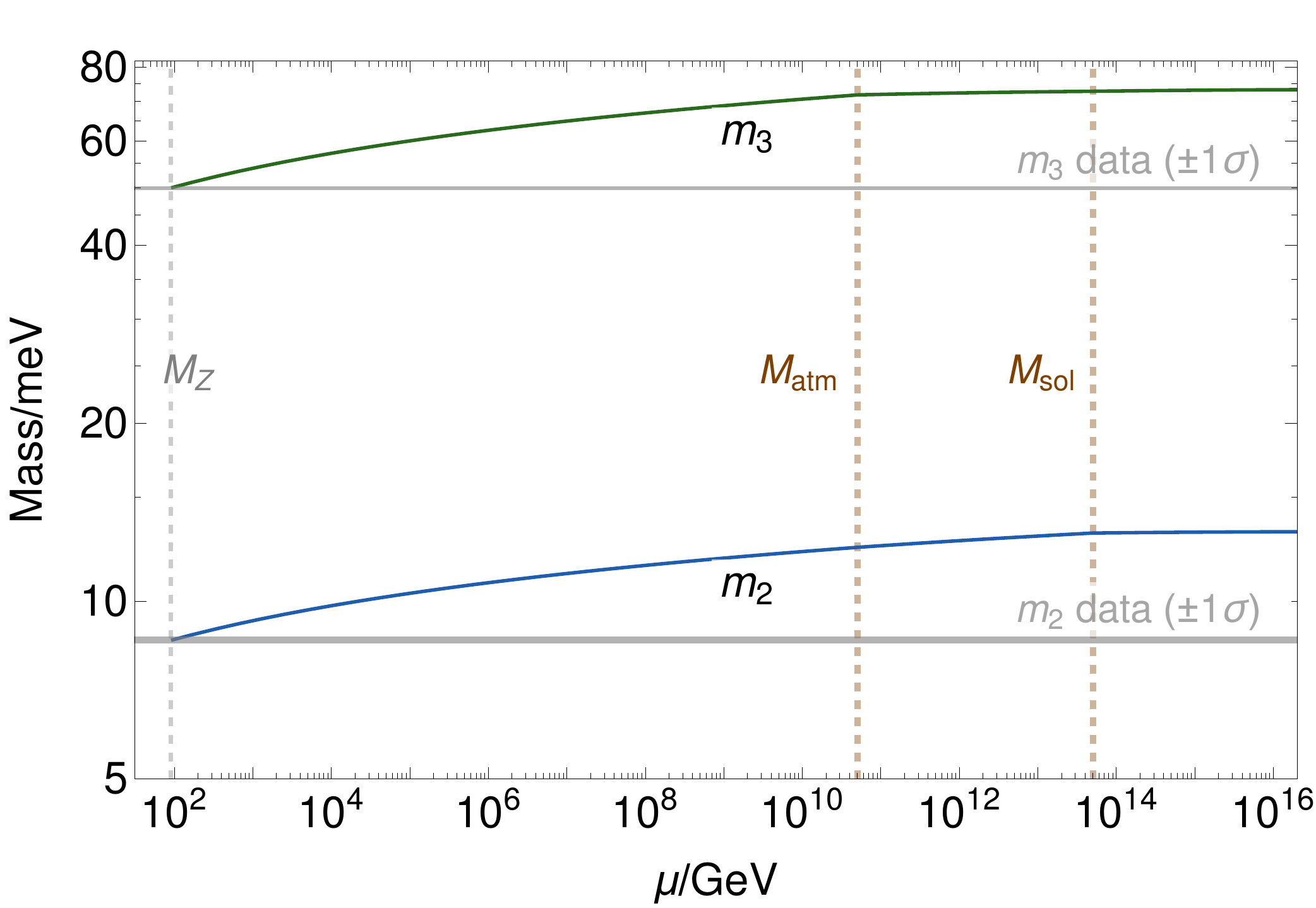}
	\includegraphics[width=0.49\textwidth]{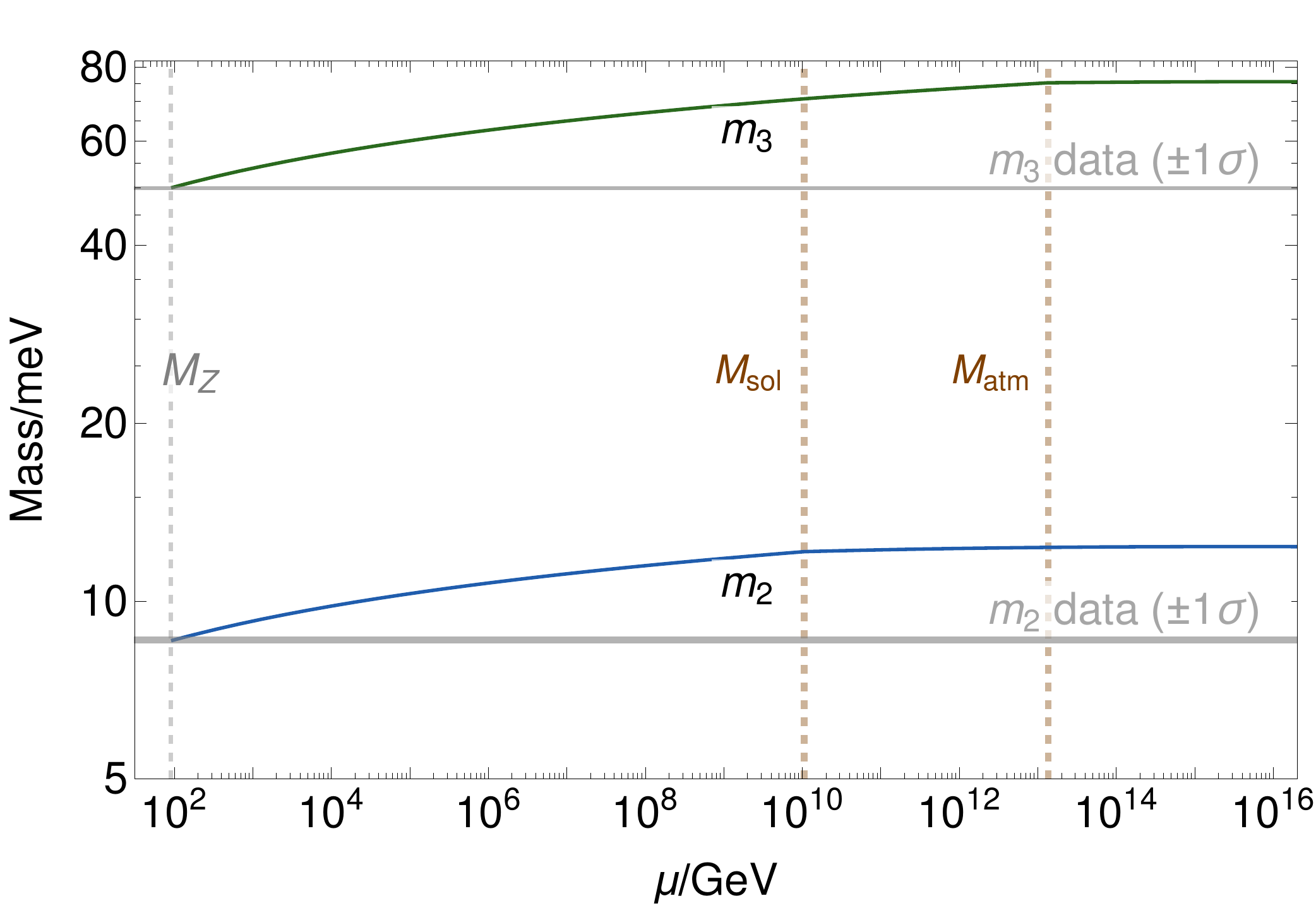}
	\includegraphics[width=0.49\textwidth]{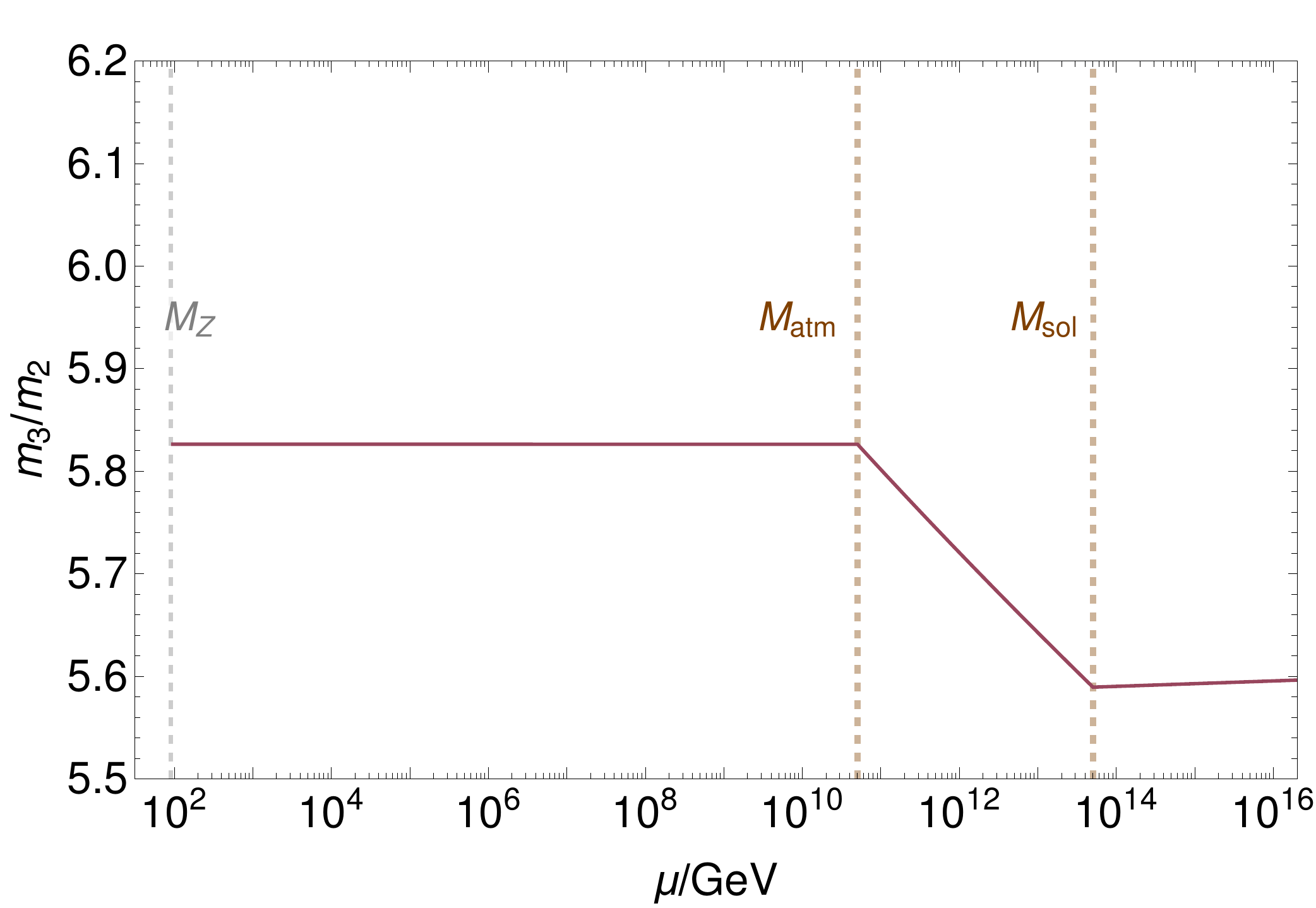}
	\includegraphics[width=0.49\textwidth]{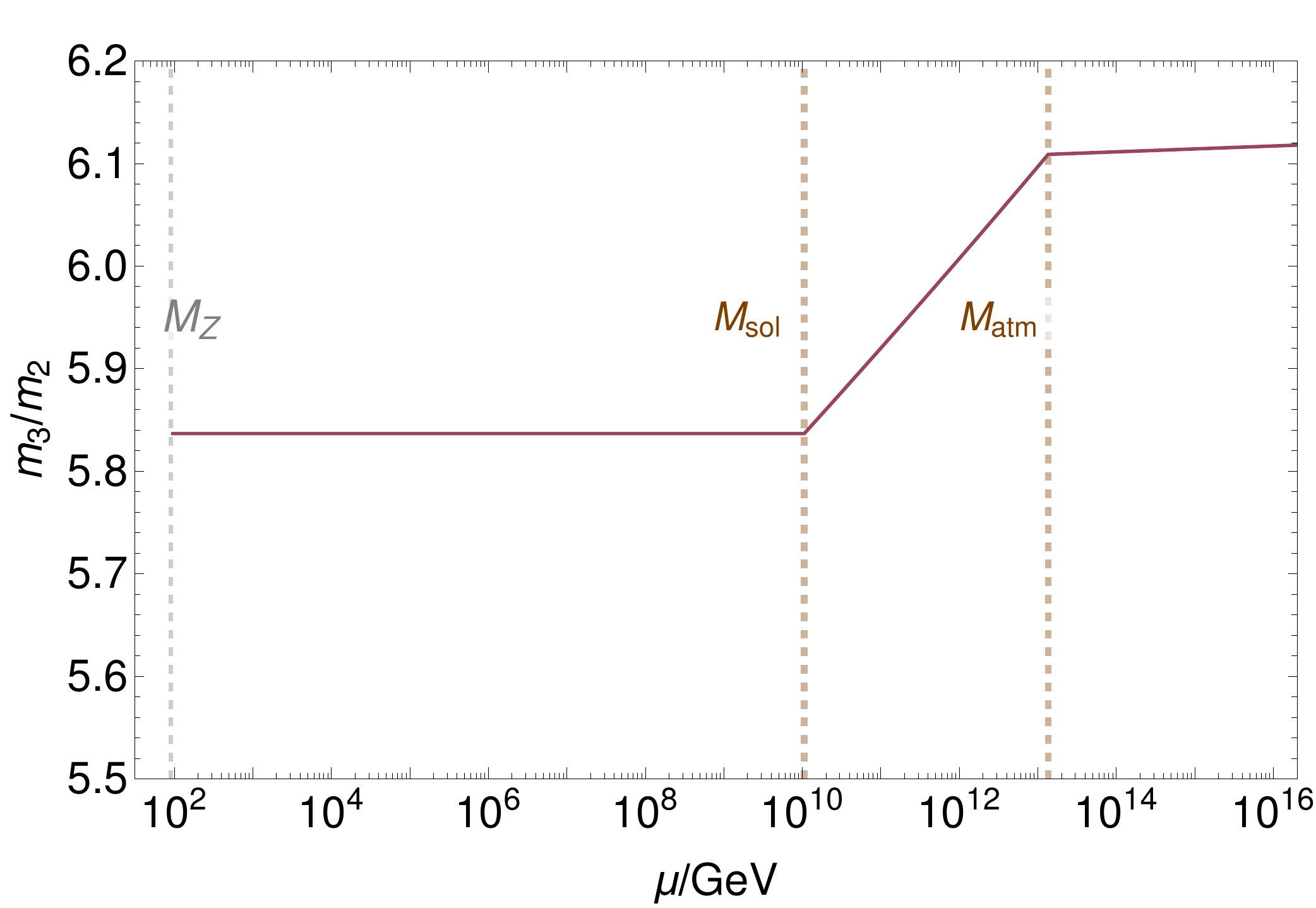}
	\caption{Running of neutrino observables. \textbf{Case A2} results are shown on the left which can be compared with \textbf{Case D2} on the right. Recalling that $m_1$=0, we have converted $1\sigma$ limits on $\Delta m_{12}^2$ and $\Delta m_{13}^2$ into limits on the $m_2$ and $m_3$ mass eigenstates.}
	\label{fig:RGE_running}
\end{figure}

We take the best fit points from \textbf{Table \ref{tab:method_2_results}} and study the \textbf{RG running} in detail, paying particular attention to the variation of neutrino mass  eigenstates and PMNS angles:

Consider \textbf{Case A2} from \textbf{Table \ref{tab:method_2_results}}. The RGE running from the GUT scale down to the electroweak (EW) scale of neutrino predictions arising from this benchmark point is presented on the left-hand side of \textbf{Figure \ref{fig:RGE_running}}. It can be seen that RGE effects on the \textbf{mass eigenstates} become very apparent below the lightest seesaw scale, when the model reduces to the SM extended by a five-dimensional Weinberg operator (see \textbf{Section \ref{RG}}).

However, for the \textbf{PMNS angles} it is a very different story. For both $\theta_{13}$ and $\theta_{23}$, RGE effects are manifest to some degree in the EFT between the two seesaw scales; however, both above and below these scales there is very little running to be seen.

In short, RGE effects are more significant in mass eigenstates than in mixing angles, but the scales at which these effects occur are vastly different between the various observables that we extract from the model. \textbf{Case D2}, shown on the right-hand side of \textbf{Figure \ref{fig:RGE_running}}, exhibits much the same behaviour as \textbf{Case A2}; large running in the mass eigenstates  and small but non-negligible running in the mixing angles.

The \textbf{ratio of masses} is also plotted for each case, to better understand the effects of running in both masses simultaneously. We see identical running for the light masses below the lowest seesaw scale for both cases, as the ratio of masses is constant below this scale. It is the running apparent between $M_2$ and $M_1$ that allows us to make a concrete prediction for the mass of the heavier RH neutrino, the lighter already being severely constrained by leptogenesis.

%

Note that the bounds on $\Delta m_{21}^2$ are not any less precise than those on $\Delta m_{31}^2$, but the mass eigenstates are shown on a logarithmic scale. There is a particularly large $1\sigma$ range for $\theta_{23}$. Discussions of future tests of the LS based on potential increases in experimental sensitivity to this observable are given in \textbf{Section \ref{sec:Future_Tests}}.

\subsection{Perturbations around Best Fit Points}

\begin{figure}[H]
\centering
\vspace{-1cm}
\begin{subfigure}[b]{\textwidth}
	\centering
	\includegraphics[width=0.495\linewidth]{{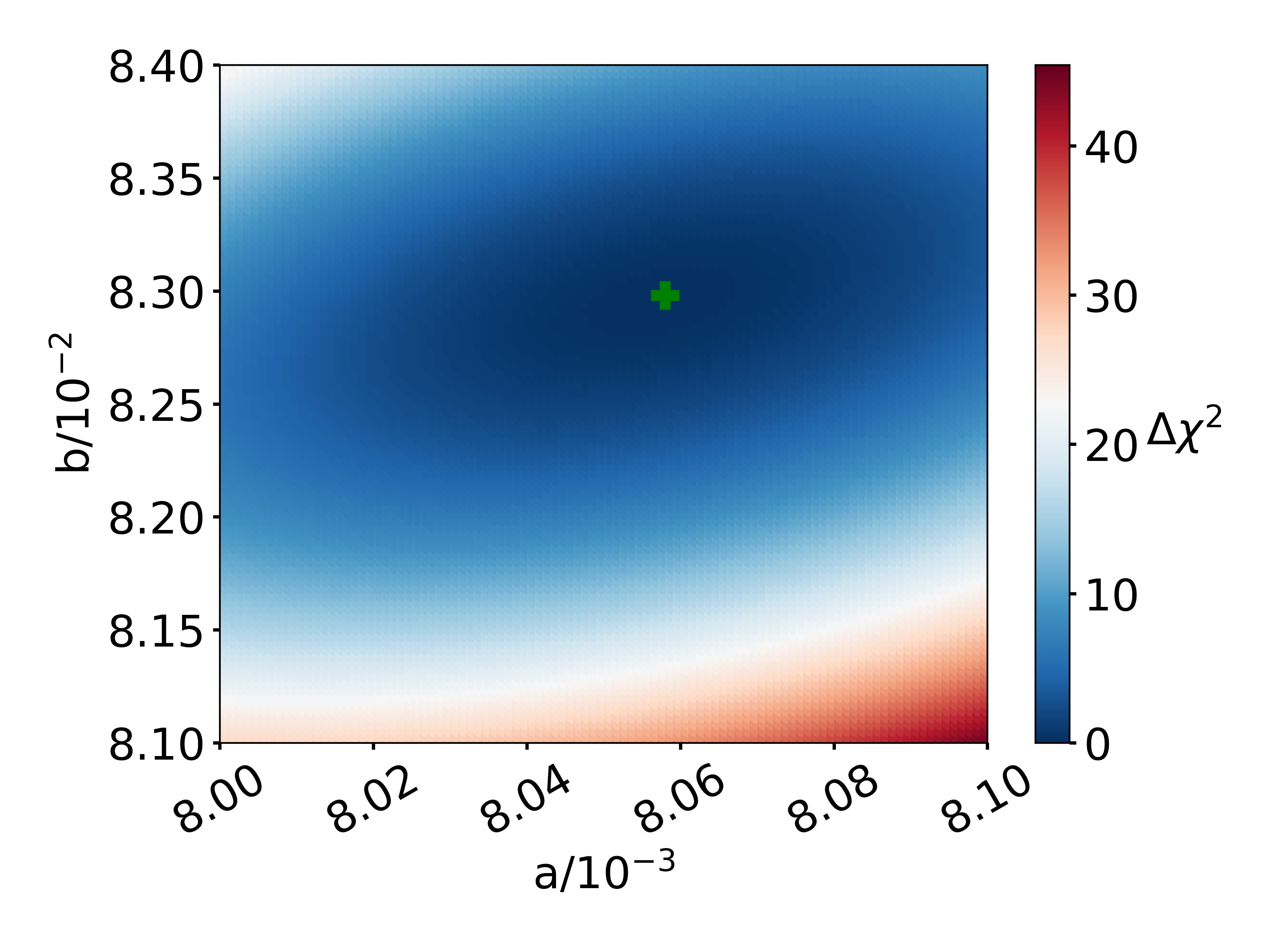}}
	\includegraphics[width=0.495\linewidth]{{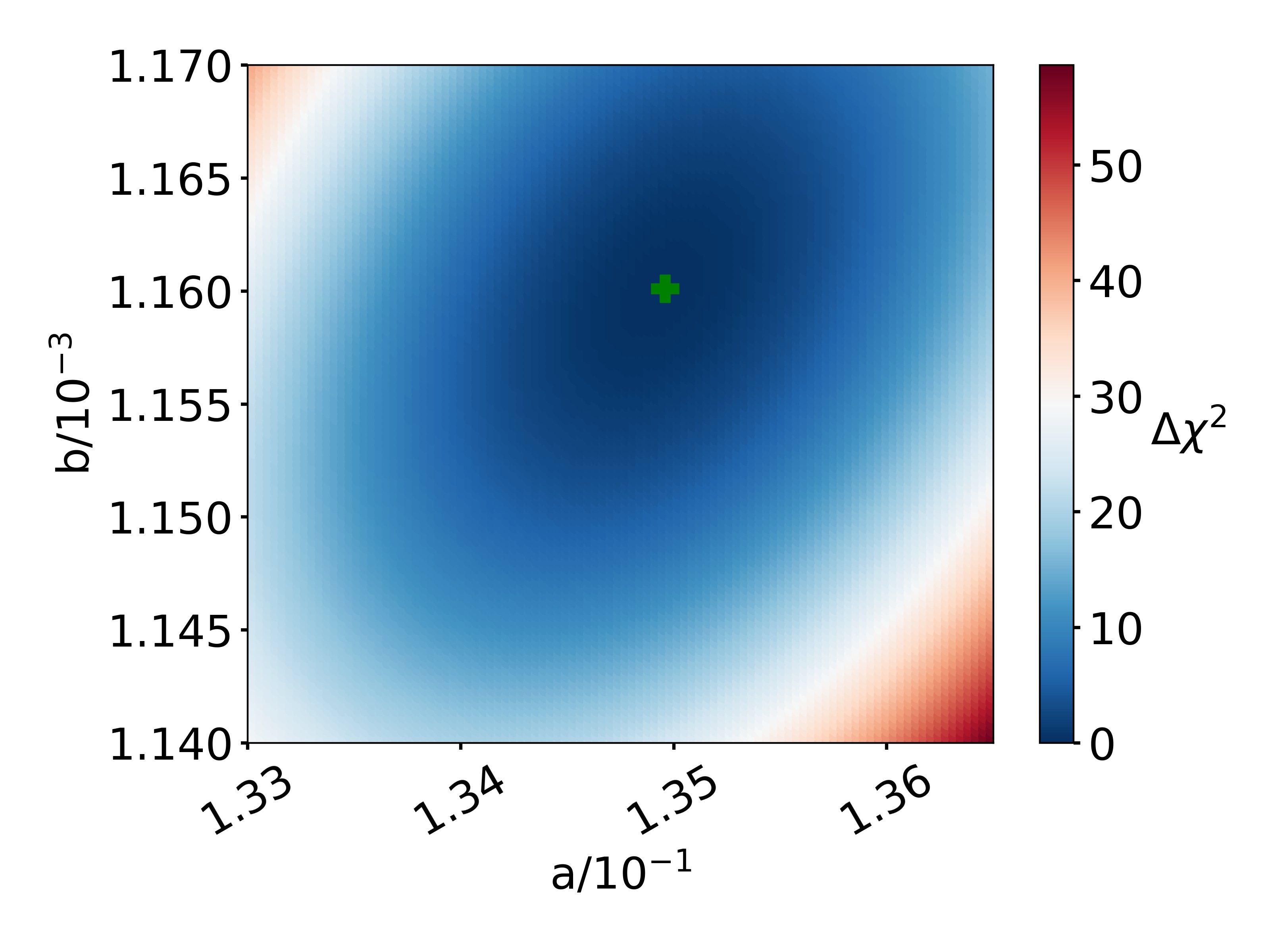}}
\end{subfigure}
\vspace{-.35cm}
\begin{subfigure}[b]{\textwidth}
	\centering
	\includegraphics[width=0.495\linewidth]{{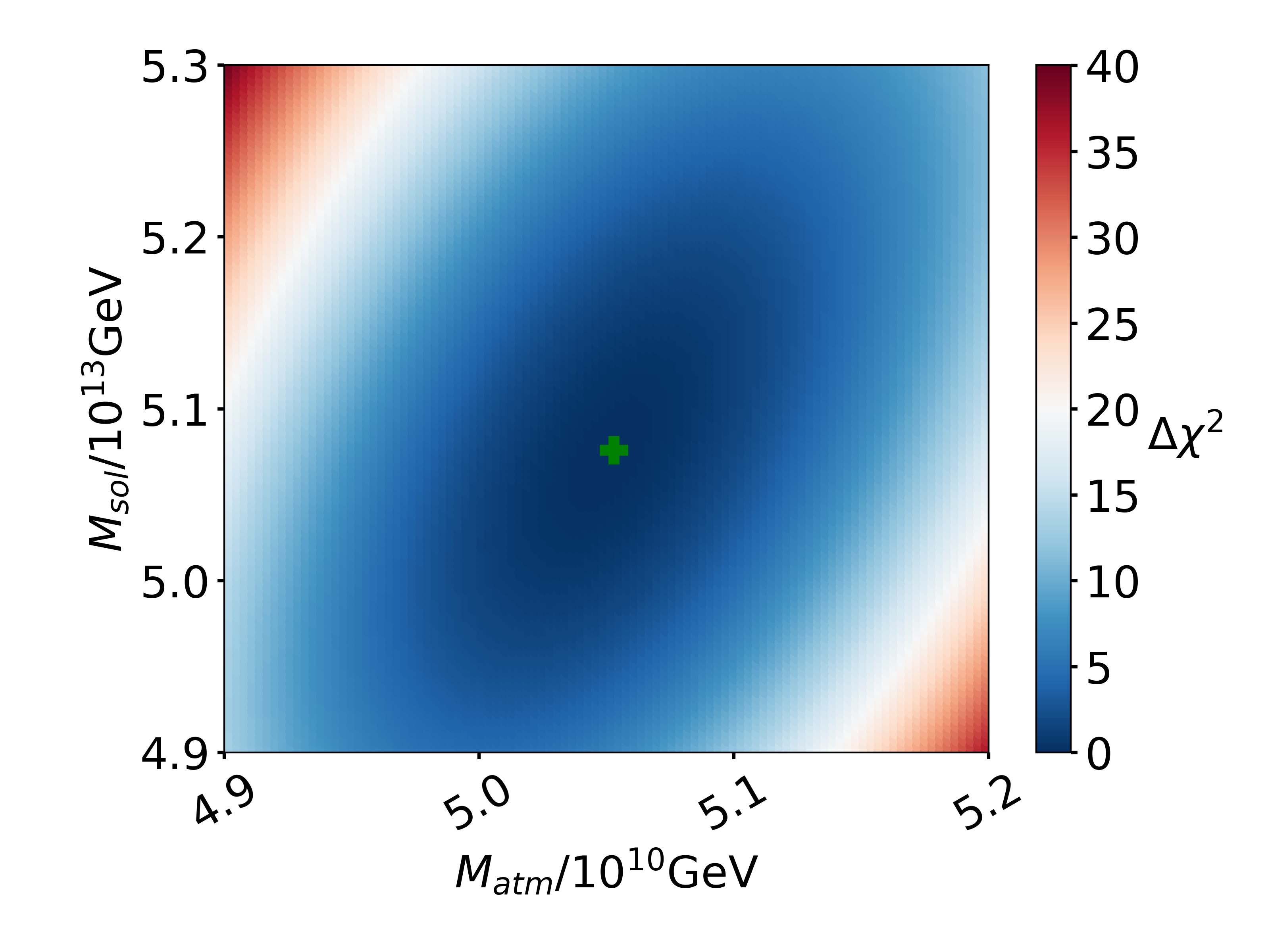}}
	\includegraphics[width=0.495\linewidth]{{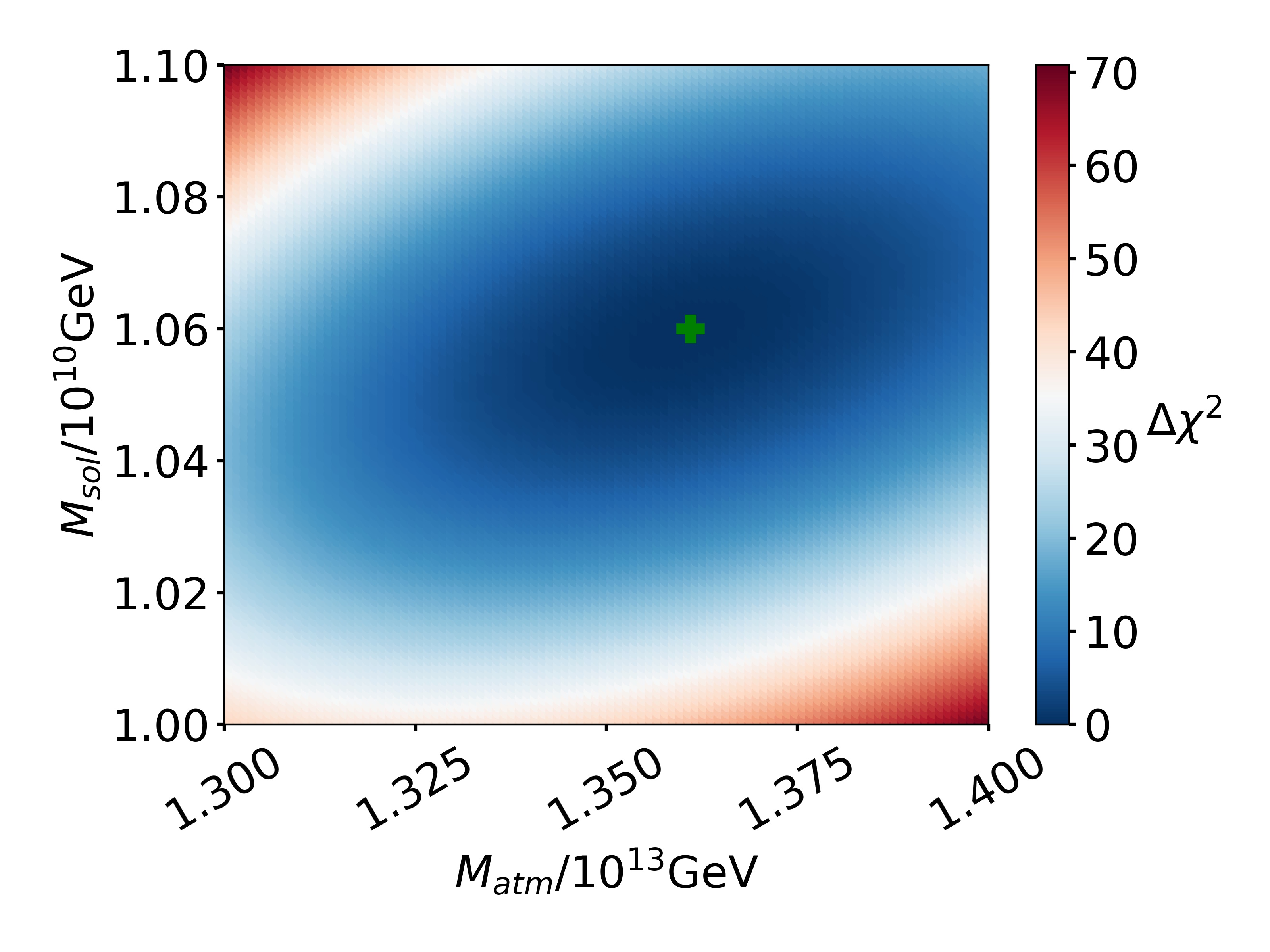}}
\end{subfigure}
\vspace{-.35cm}
\caption{Perturbations around \textbf{Case A2} benchmark point shown on the left, those for \textbf{Case D2} on the right. In each case, two parameters are varied at a time while the other two are kept fixed. Differently coloured circles represent approximate 1, 2 and 3 sigma deviations from the best fit in each parameter, and the green cross marks the benchmark point.}
\label{fig:heatmaps}
\end{figure}

It is useful to show the best fit points we obtain with this analysis visually (see \textbf{Table \ref{tab:method_2_results}} for their numerical values). In this section, we vary our input parameters around these benchmark points in both one and two dimensions, and we see that such \textbf{perturbations in parameter space} yield variations around smooth, stable minima. \textbf{Figure \ref{fig:heatmaps}} shows heat maps representing increases in $\chi^2$ as one moves away from the benchmark points, for variations in $a,b$ or $M_{atm},M_{sol}$ parameter space, respectively. Note the resulting shape is never an exact circle, as the analysis is not sensitive to all parameters equally. \\ \\ \\

\begin{figure}[H]
\centering
\begin{subfigure}[b]{\textwidth}
	\centering
	\includegraphics[width=0.45\linewidth]{{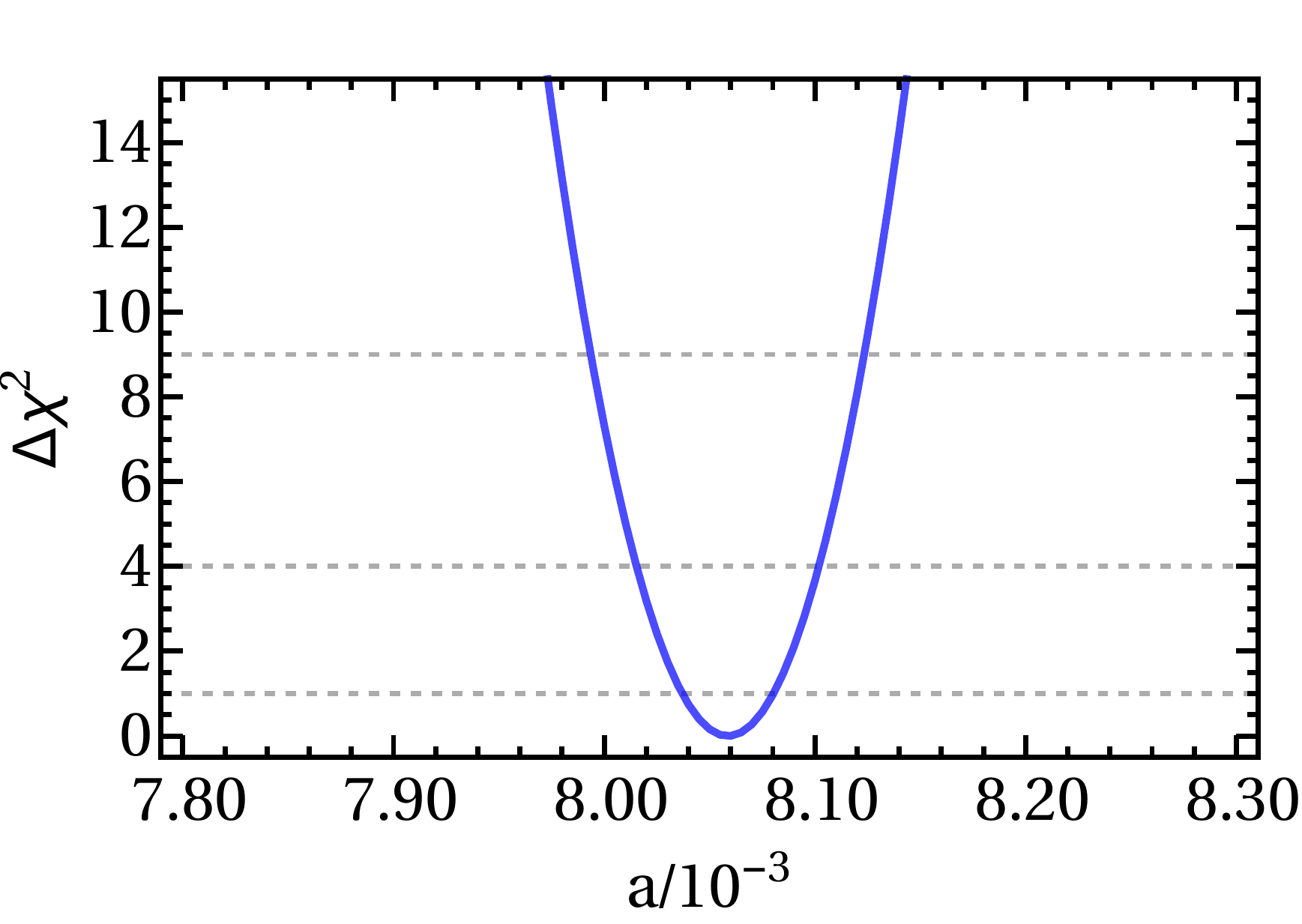}}
	\includegraphics[width=0.45\linewidth]{{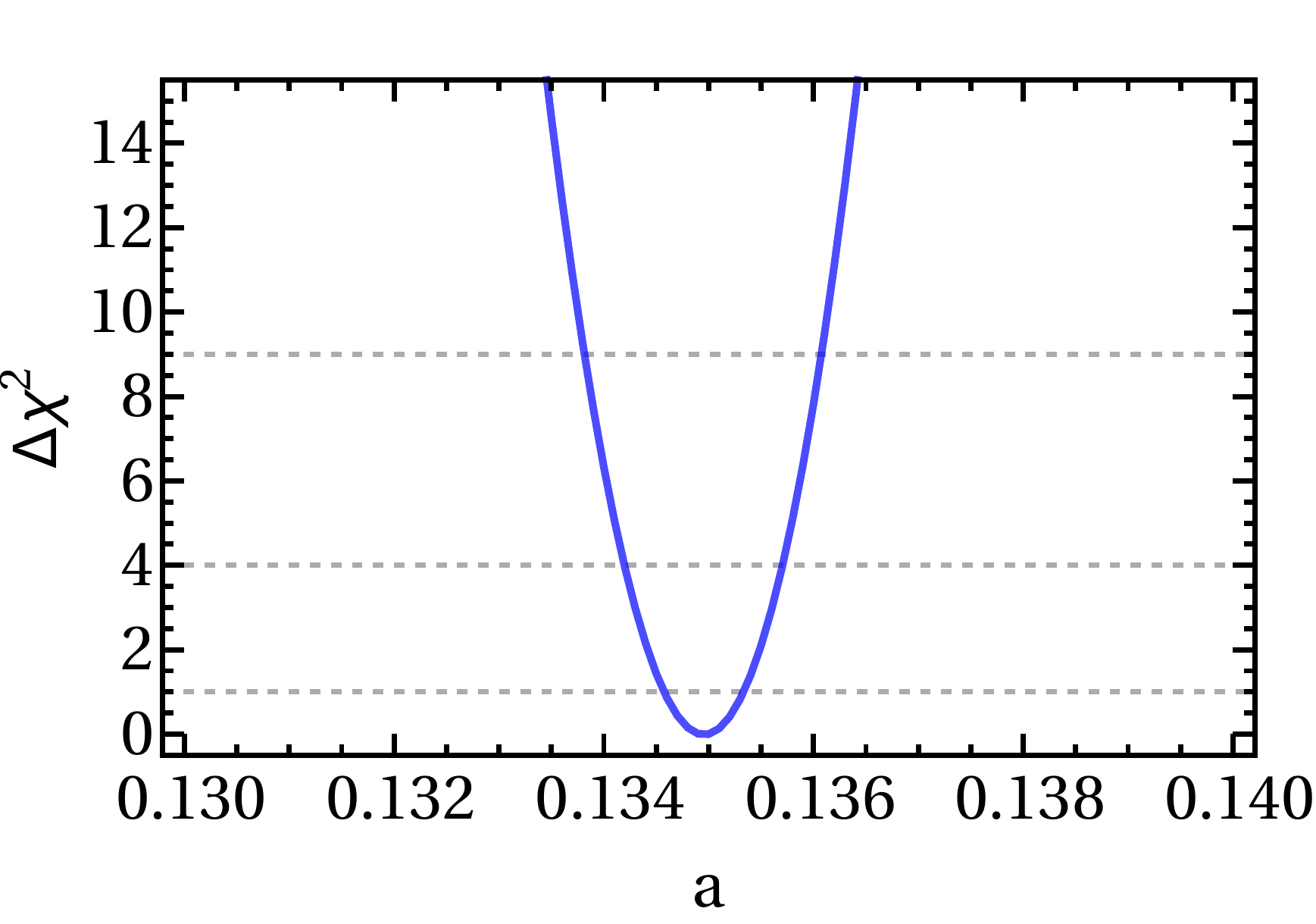}}
\end{subfigure}
\vspace{-.15cm}
\begin{subfigure}[b]{\textwidth}
	\centering
	\includegraphics[width=0.45\linewidth]{{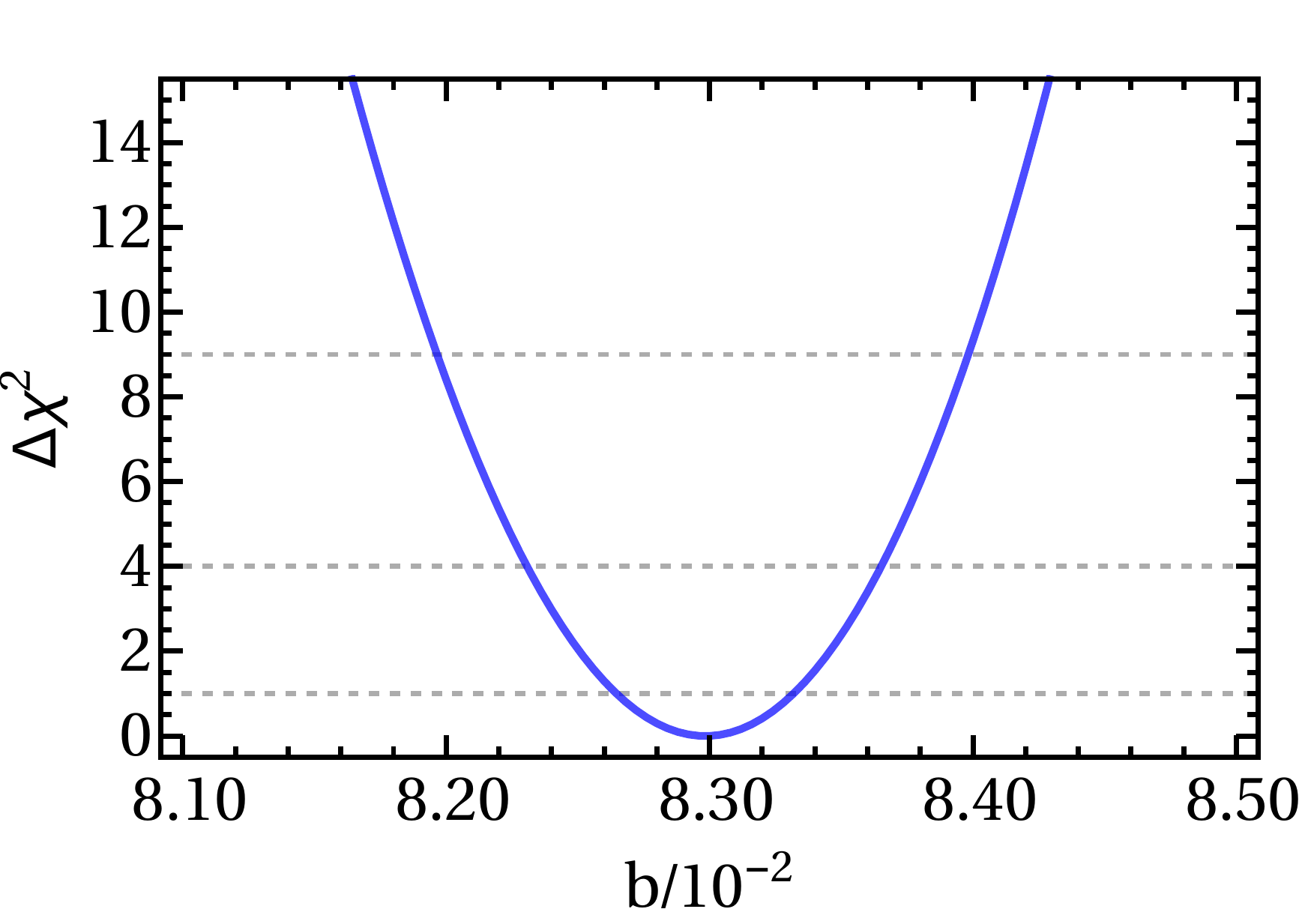}}
	\includegraphics[width=0.45\linewidth]{{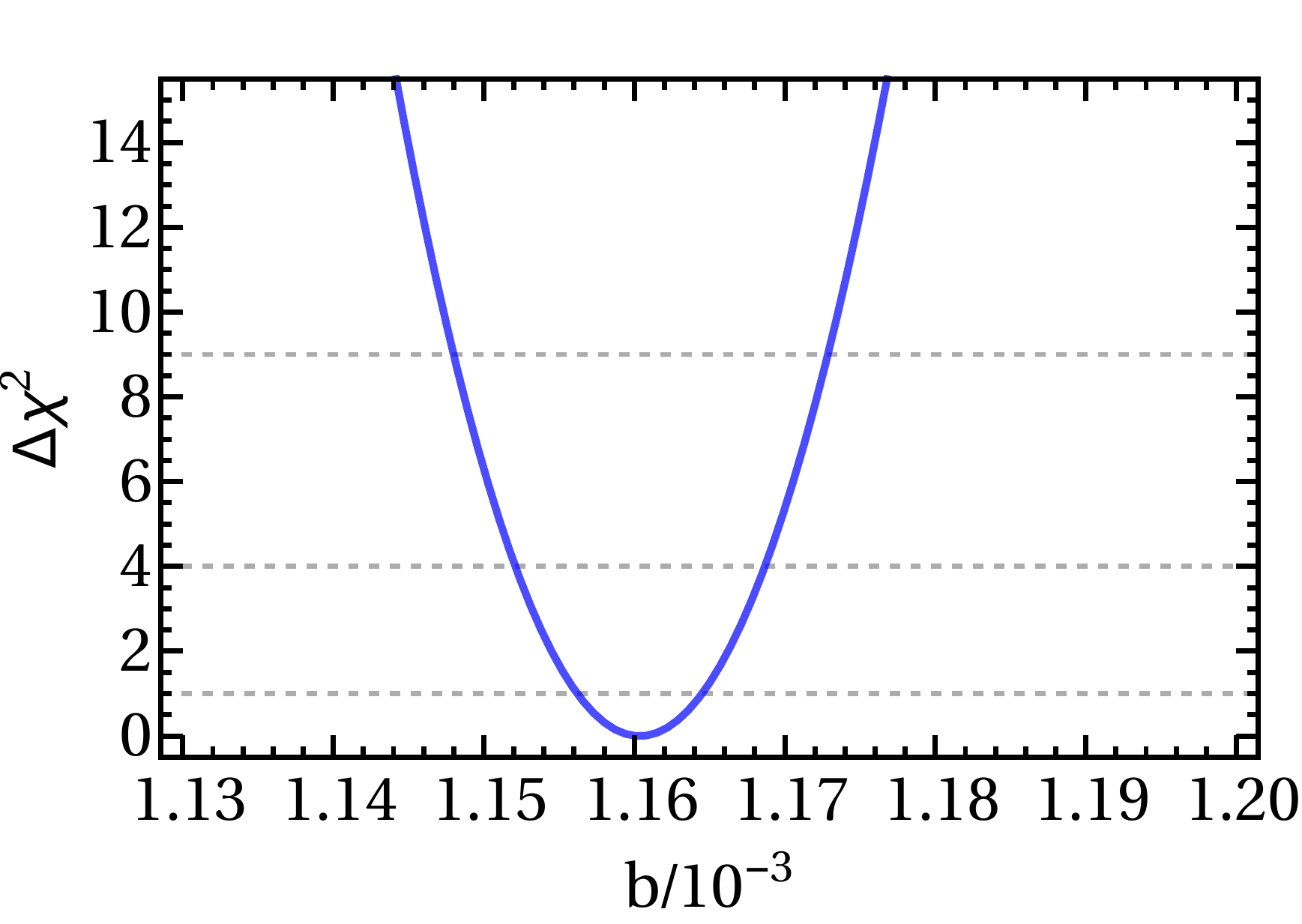}}
\end{subfigure}
\vspace{-.15cm}
\begin{subfigure}[b]{\textwidth}
	\centering
	\includegraphics[width=0.45\linewidth]{{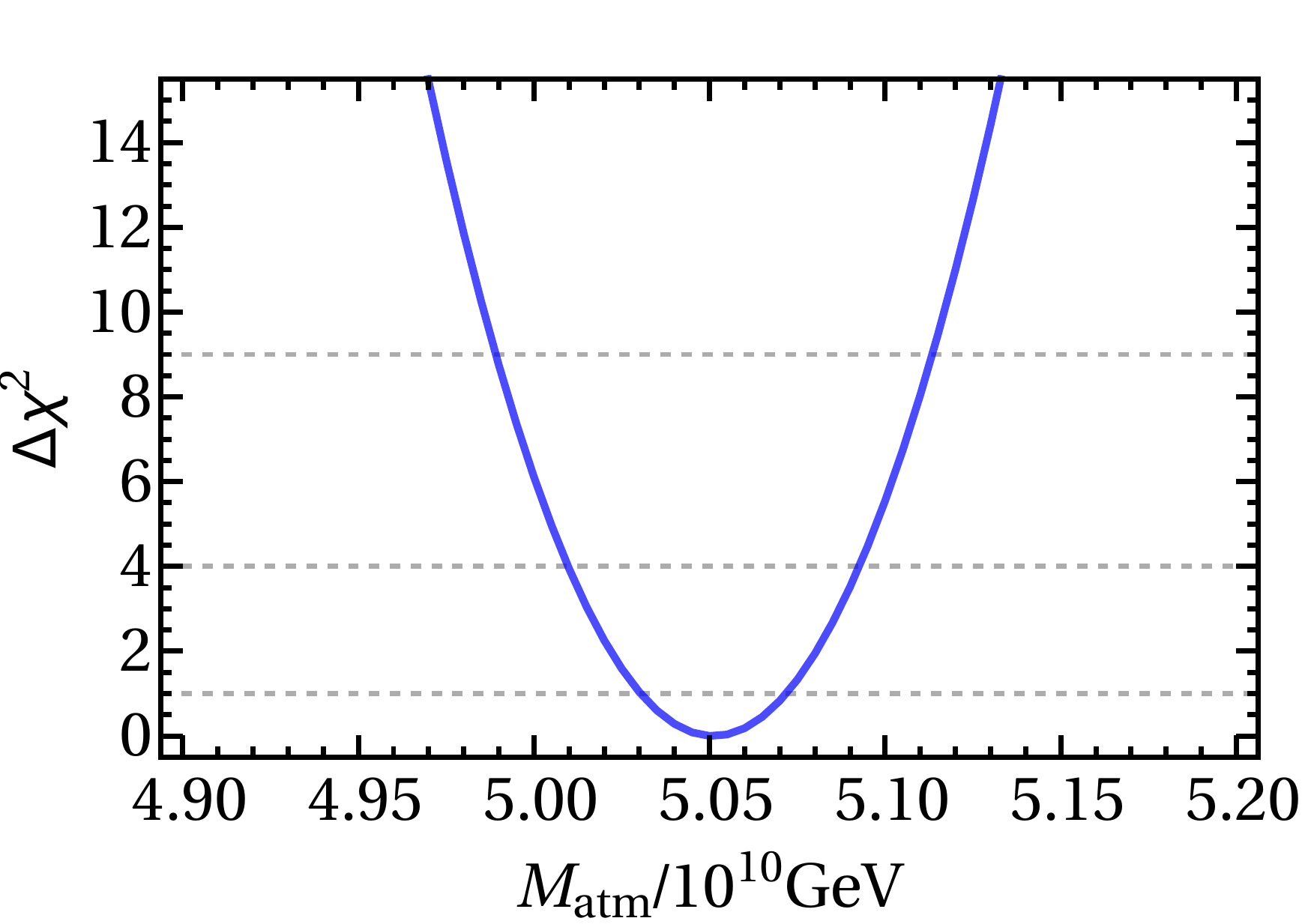}}
	\includegraphics[width=0.45\linewidth]{{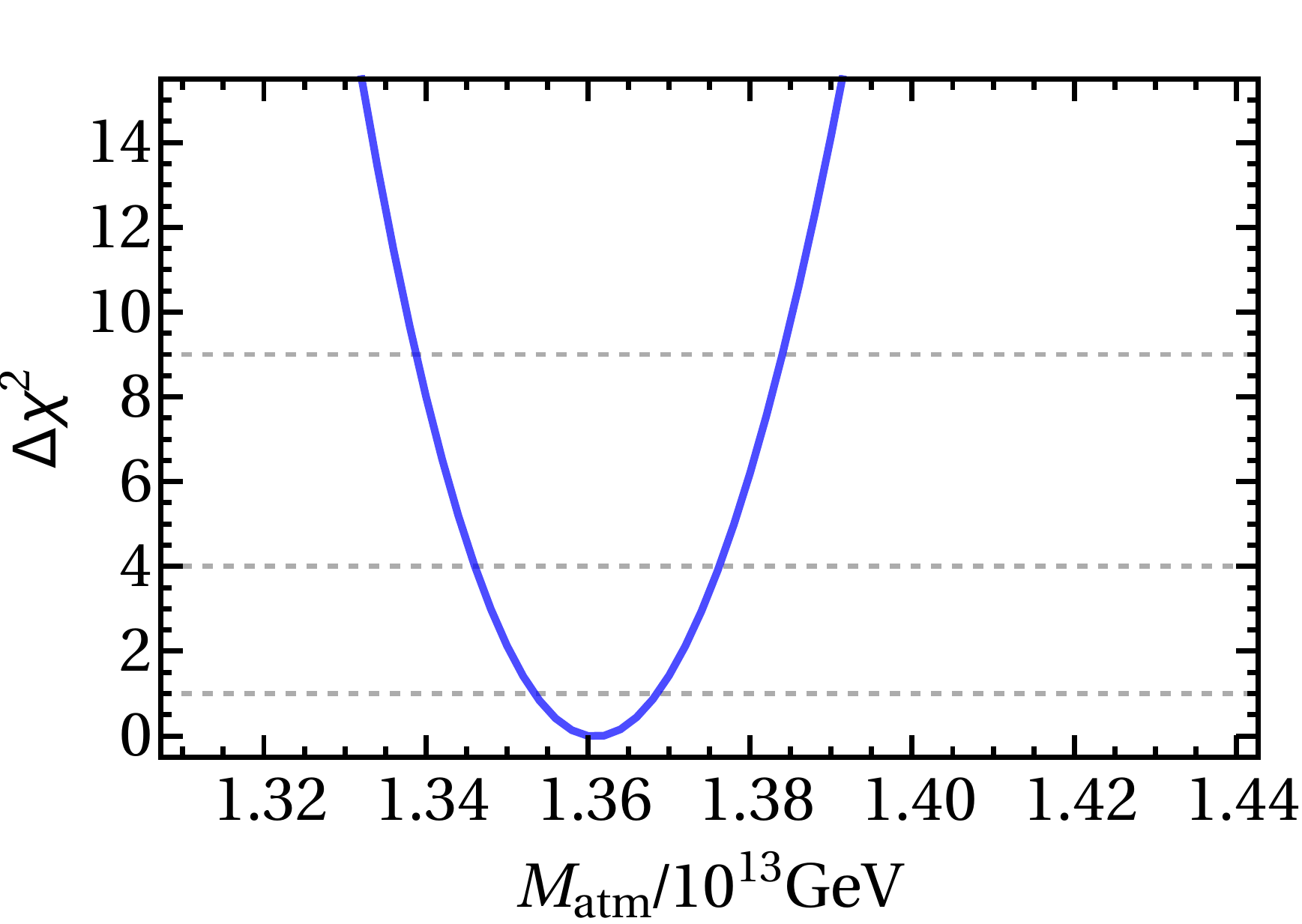}}
\end{subfigure}
\vspace{-.15cm}
\begin{subfigure}[b]{\textwidth}
	\centering
	\includegraphics[width=0.45\linewidth]{{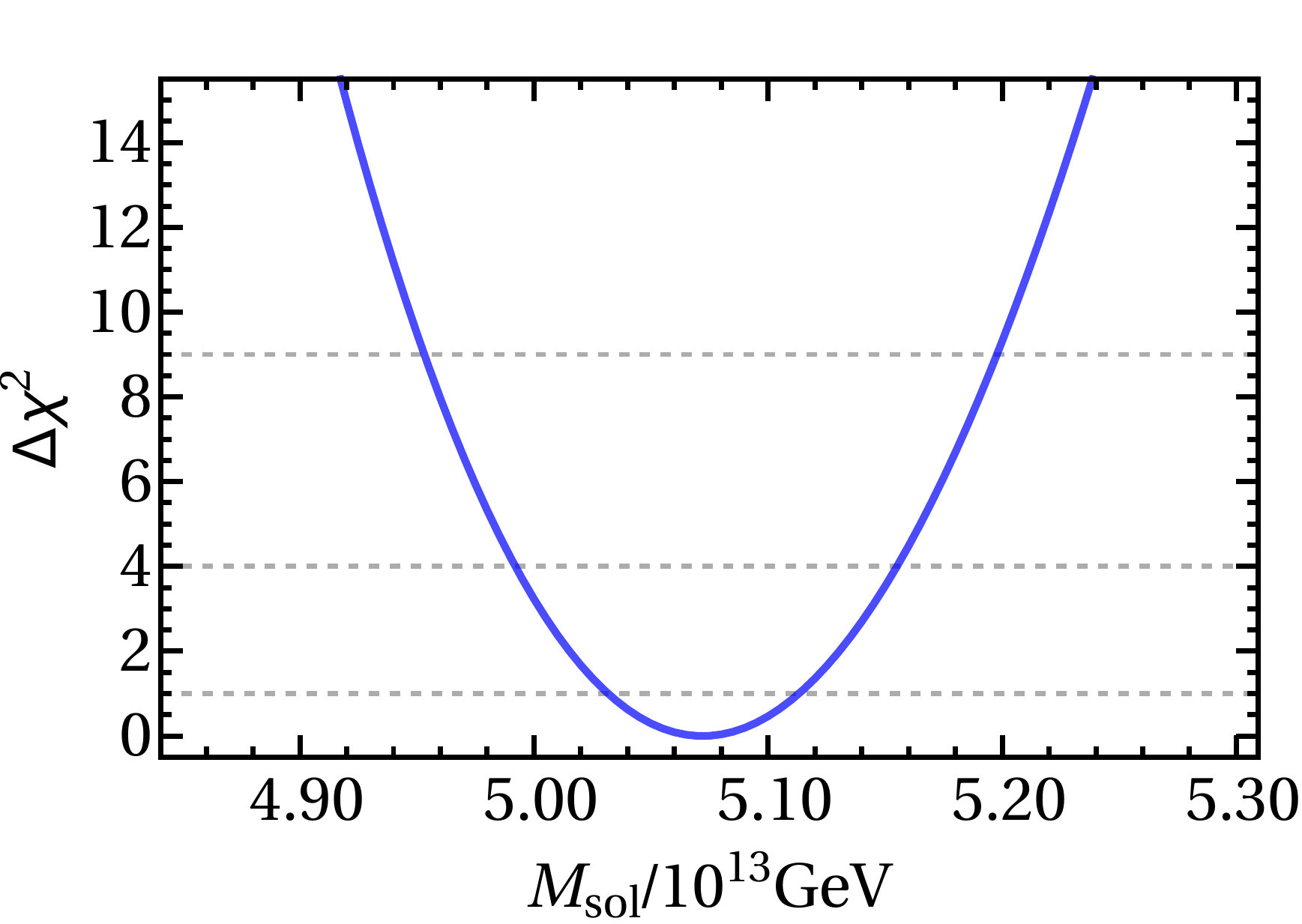}}
	\includegraphics[width=0.45\linewidth]{{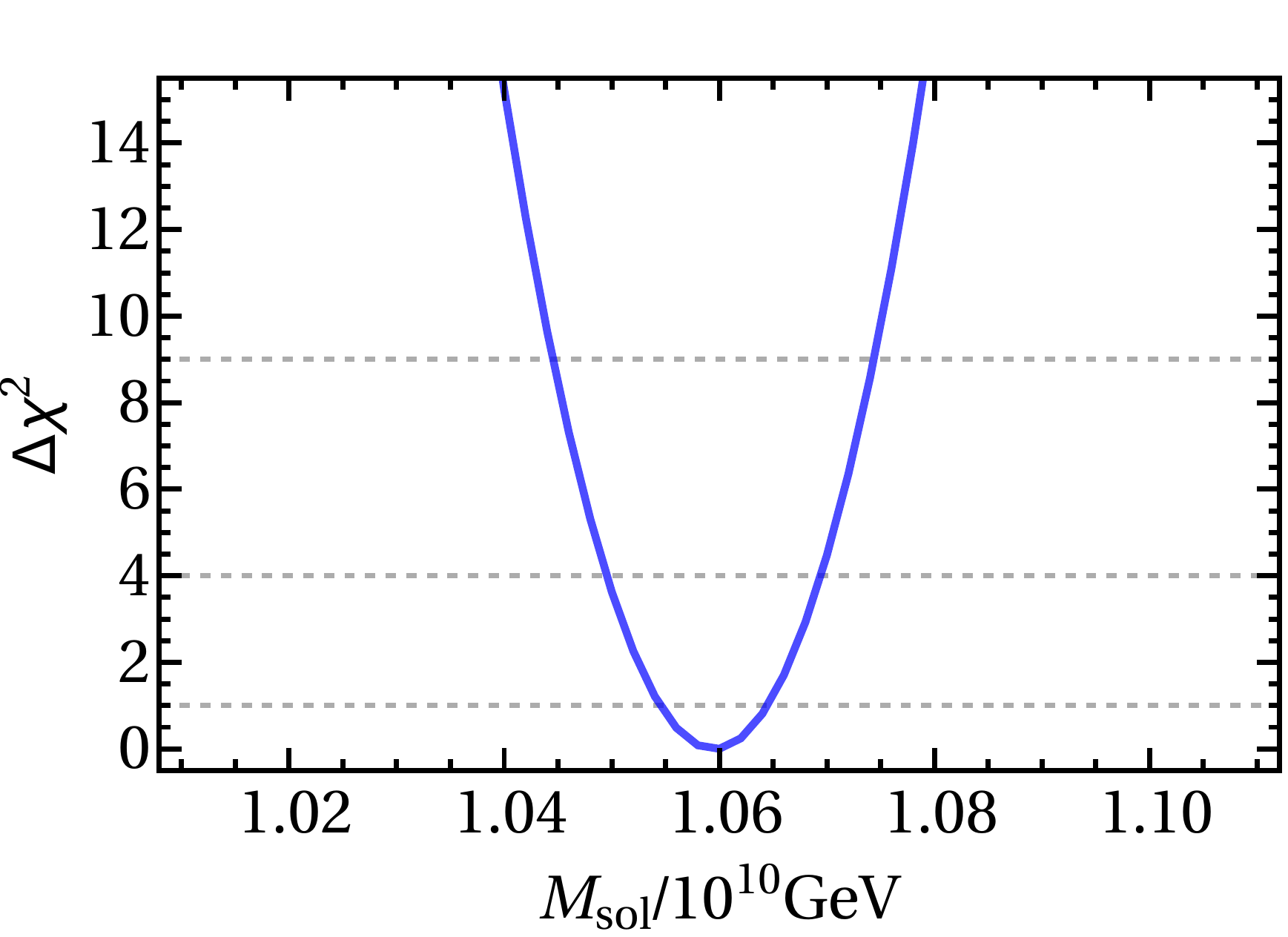}}
\end{subfigure}
\vspace{-.15cm}
\caption{Perturbations around \textbf{Case A2} benchmark point shown on the left, those for \textbf{Case D2} on the right. Each parameter is varied separately as the other three are kept fixed. Dashed gray lines show 1, 2 and 3 sigma deviations from the best fit in each parameter when varied separately.}
\label{fig:caseA}
\end{figure}

We now vary each parameter individually around the best fit points given in \textbf{Cases A2} and \textbf{D2}, whilst keeping the other three parameters fixed - \textbf{Figure \ref{fig:caseA}} shows such perturbations. On the vertical axes, $\Delta\chi^2$ is the deviation from minimum $\chi^2$; the stationary point thus shows a vanishing $\Delta\chi^2$ corresponding to the benchmark point itself.

\subsection{Future Tests of the Littlest Seesaw}
\label{sec:Future_Tests}
Given the constantly evolving nature of particle physics and the rapid technological advances being made in neutrino experiments, it is to be expected that the precision of PMNS parameter measurements will improve considerably in the coming years. With this in mind, it seems pertinent to discuss the \textbf{range of values} of each observable for which this analysis method of the Littlest Seesaw model remains a relevant and viable test of neutrino masses and properties.

\textbf{Table \ref{tab:A-observables-ranges}} below shows $1\,\sigma$, $2\,\sigma$ and $3\,\sigma$ ranges for each of the observables predicted by the Littlest Seesaw model in our analysis of \textbf{Case A2}. 


\begin{table}[H]
	\centering
	\begin{tabular}{lrrr}
	\toprule \addlinespace
	& \textbf{$1\,\sigma$ range} & \textbf{$2\,\sigma$ range} & \textbf{$3\,\sigma$ range}\\ \addlinespace
	\midrule \addlinespace
	$\theta_{12}/^{\circ}$ & $34.254 \rightarrow 34.350$ & $34.236 \rightarrow 34.365$ & $34.217 \rightarrow 34.383$\\ \addlinespace
	$\theta_{13}/^{\circ}$ & $8.370 \rightarrow 8.803$ & $8.300 \rightarrow 8.878$ & $8.218 \rightarrow 8.959$\\ \addlinespace
	$\theta_{23}/^{\circ}$ & $45.405 \rightarrow 45.834$ & $45.343 \rightarrow 45.910$ & $45.269 \rightarrow 45.996$\\ \addlinespace
	$\Delta {m_{12}}^2/10^{-5}\text{eV}^2$ & $7.030 \rightarrow 7.673$ & $6.930 \rightarrow 7.805$ & $6.788 \rightarrow 7.952$\\ \addlinespace
	$\Delta {m_{31}}^2/10^{-3}\text{eV}^2$ & $2.434 \rightarrow 2.561$ & $2.407 \rightarrow 2.587$ & $2.377 \rightarrow 2.616$\\ \addlinespace
	$\delta/^{\circ}$ & $- 88.284 \rightarrow - 86.568$ & $- 88.546 \rightarrow - 86.287$ & $-88.864 \rightarrow -85.966$\\ \addlinespace
	$Y_B/10^{-10}$ & $0.839 \rightarrow 0.881$ & $0.831 \rightarrow 0.889$ & $0.822 \rightarrow 0.898$\\ \addlinespace
	\bottomrule \addlinespace
	\end{tabular}
	\caption{Ranges of observables for \textbf{Case A2}.}
	\label{tab:A-observables-ranges}
\end{table}

The same ranges are shown for \textbf{Case D2} in \textbf{Table \ref{tab:D-observables-ranges}}. It is interesting to note that for \textbf{Case D}, the values of $\theta_{23}$ favoured by the model are slightly lower than in \textbf{Case A}, as are the predicted values of $\delta$.

\begin{table}[H]
	\centering
	\begin{tabular}{lrrr}
	\toprule \addlinespace
	& \textbf{$1\,\sigma$ range} & \textbf{$2\,\sigma$ range} & \textbf{$3\,\sigma$ range}\\ \addlinespace
	\midrule \addlinespace
	$\theta_{12}/^{\circ}$ & $34.291 \rightarrow 34.379$ & $34.278 \rightarrow 34.391$ & $34.264 \rightarrow 34.404$\\ \addlinespace
	$\theta_{13}/^{\circ}$ & $8.384 \rightarrow 8.784$ & $8.329 \rightarrow 8.838$ & $8.268 \rightarrow 8.902$\\ \addlinespace
	$\theta_{23}/^{\circ}$ & $44.044 \rightarrow 44.434$ & $43.991 \rightarrow 44.484$ & $43.925 \rightarrow 44.539$\\ \addlinespace
	$\Delta {m_{12}}^2/10^{-5}\text{eV}^2$ & $7.058 \rightarrow 7.615$ & $6.966 \rightarrow 7.688$ & $6.875 \rightarrow 7.787$\\ \addlinespace
	$\Delta {m_{31}}^2/10^{-3}\text{eV}^2$ & $2.435 \rightarrow 2.562$ & $2.407 \rightarrow 2.590$ & $2.373 \rightarrow 2.624$\\ \addlinespace
	$\delta/^{\circ}$ & $- 93.708 \rightarrow - 92.180$ & $- 93.919 \rightarrow - 91.964$ & $- 94.160 \rightarrow - 91.730$\\ \addlinespace
	$Y_B/10^{-10}$ & $0.838 \rightarrow 0.881$ & $0.827 \rightarrow 0.893$ & $0.820 \rightarrow 0.899$\\ \addlinespace
	\bottomrule \addlinespace
	\end{tabular}
	\caption{\textbf{Case D2} ranges for observables}
	\label{tab:D-observables-ranges}
\end{table}

In other words, if future neutrino experiments were to precisely measure these values as well outside these ranges, this would be a way to disqualify this method of testing the LS and perhaps even guide theoretical models towards a different understanding of the characteristics of neutrinos.

This statement is of particular relevance when it comes to $\theta_{23}$, as it can be seen that any experimentally measured departure from close to maximal mixing would come in direct contradiction with one of the intrinsic features of the model. It is also interesting to note once again the case of $\delta$ - although the experimental uncertainty on this parameter is at present extremely large, the fits performed in this analysis provided a suggestion for $\delta\sim-90^\circ$, which is consistent with the newest experimental hints.

However, the ranges stated above are purely indicative and should not be taken as absolute. The method used to obtain them incurs limitations, as it was not possible to analyse the range of each observable separately, due to the way the analysis was set up. The $\chi^2$ values used to define the $\sigma$ ranges studied were made up of contributions from all seven observables at the same time, which explains why the final valid ranges of observables are rather narrow (as the effect of, for instance, $Y_B$ will be more dominant on the total $\chi^2$ than that of one of the mixing angles).

As an additional test of the model, therefore, we hypothesised a possible future experimental sensitivity on $\theta_{23}$, based on the work of \cite{Ballett:2016yod}. Taking the current measured central value of $\theta_{23}=47.2 \degree$ with its expected future $1\sigma$ precision of $\pm\,0.66 \degree$ - based on a combined sensitivity analysis of DUNE and T2HK - we perform a new scan to see whether this measurement would greatly alter the $\chi^2$ and thus invalidate the LS. The results for both \textbf{Case A} and \textbf{D} - denoted by \textbf{A2'} and \textbf{D2'} respectively - are shown in \textbf{Table \ref{tab:future_theta23}} and compared to the results obtained previously. \\


\begin{table}[H]
	\centering
	\begin{tabular}{lrrrr}
	\toprule \addlinespace
	& \textbf{Case A2} & \textbf{Case A2'} & \textbf{Case D2} & \textbf{Case D2'} \\ \addlinespace
	\midrule \addlinespace
	$\chi^2$/d.o.f. & $1.749/3$ & $7.49/3$ & $2.070/3$ & $21.80/3$ \\ \addlinespace
	\bottomrule \addlinespace
	\end{tabular}
	\caption{Best Fit Point $\chi^2$ for a hypothetical future $\theta_{23}$ measurement. Corresponding best fit parameters shown in \textbf{Table \ref{tab:method_2_results}}.}
	\label{tab:future_theta23}
\end{table}
It can be seen that the $\chi^2$ of \textbf{Case D2} suffers a very large increase for this new hypothetical experimental value of $\theta_{23}$, which would seem to rule out this case as a viable test of the model. However, \textbf{Case A2} would only see its $\chi^2$ pushed to a value of $2.49$ per d.o.f., which could possibly be improved with additional testing. We can therefore conclude that if this observable were indeed measured very precisely to be its current value, the LS could potentially continue to be a valid model, in spite of a departure from close to maximal mixing.

\section{Conclusions}
\label{conclusion}

The Littlest Seesaw (LS) model remains the most minimal seesaw model which can explain current data 
with the smallest number of parameters. It involves two right-handed neutrino masses plus two
real coefficients $a,b$ of the column vectors proportional to $(0,1,1)$ and $(1,3,1)$ or $(1,1,3)$,
comprising the Yukawa matrix in the flavour basis, with a fixed relative phase between these columns of $\mp \pi/3$.
In this paper we have performed the first global fit of the LS model to low-energy neutrino data
and leptogenesis, taking into account RG corrections.
We have shown that the four high-energy LS parameters in the flavour basis, namely two real
Yukawa couplings $a,b$
plus the two right-handed neutrino masses, can be determined by an excellent fit to the seven currently constrained observables of low-energy neutrino data and leptogenesis. 

For \textbf{Case A} in Eq.~\ref{eq:YnuA}, corresponding to Yukawa columns $a(0,1,1)$ and $b(1,3,1)$, we fit the respective right-handed neutrino
masses to be $M_{atm}\approx 5\times 10^{10}$ GeV and $M_{sol}\approx 0.5-3\times 10^{14}$ GeV, depending 
on the GUT scale.
For \textbf{Case D} in Eq.~\ref{eq:YnuD}, corresponding to Yukawa columns $b(1,1,3)$ and $a(0,1,1)$, we fit the respective right-handed neutrino
masses to be $M_{sol}\approx 1\times 10^{10}$ GeV and $M_{atm}\approx 1.6-14\times 10^{12}$ GeV, depending 
on the GUT scale.
We estimate $\chi^2 \simeq 1.5-1.75$ for the three d.o.f. for \textbf{Case A},
and $\chi^2 \simeq 2.1-2.6$ for the three d.o.f. for \textbf{Case D}, depending 
on the GUT scale. Both are excellent fits, regardless of the assumed unification scale.
We extract allowed ranges of neutrino parameters from our fit data, including $\theta_{23}=45.3^o-46.0^o$
and $\delta = -87^o\pm 2^o$ for \textbf{Case A} and $\theta_{23}=44.0^o-44.5^o$
and $\delta = -93^o\pm 2^o$ for \textbf{Case D}.
Both cases predict normal mass ordering with $m_1=0$.
These results will enable LS models to be tested in future neutrino experiments.

In conclusion, the Littlest Seesaw continues to be a relevant, highly consistent model that provides an outstanding fit to data. It is predictive, with just four high-energy parameters resulting in seven observables apparent at low scales. 
Taking into account RG corrections, 
this enables both the high-energy RHN masses and the Yukawa coupling constants $a,b$
to be fixed by low-energy neutrino data and leptogenesis, for the first time in any seesaw model.
In turn, the resulting fit gives restricted ranges of low-energy observables, where these predictions 
will be confronted by future neutrino experiments.
Within this framework, future neutrino experiments will allow a window into the 
GUT scale parameters of the most minimal seesaw model, providing 
insight into physics at the highest scales.


\pagebreak

\subsection*{Acknowledgements}
The authors would like to thank Fredrik Bj\"orkeroth, Pasquale Di Bari, Tanja Geib and Nick Prouse for their input in useful discussions.
S.\,F.\,K. acknowledges the STFC Consolidated Grant ST/L000296/1 and the European Union's Horizon 2020 Research and Innovation programme under Marie Sk\l{}odowska-Curie grant agreements Elusives ITN No.\ 674896 and InvisiblesPlus RISE No.\ 690575.
S.\,M.\,S. wishes to acknowledge the STFC Doctoral Training Grant held by Queen Mary University of London and additional support from the University of Southampton and the 
Valerie Myerscough Science and Mathematics Trust Fund at the University of London.
S.\,J.\,R. acknowledges support from a Mayflower PhD studentship at the University of Southampton.


\begin{thebibliography}{99}
	\setlength{\itemsep}{0em}
	
	  \bibitem{nobel}
	 Special Issue on
	 ``Neutrino Oscillations: Celebrating the Nobel Prize in Physics 2015''
	 Edited by Tommy Ohlsson,
	 Nucl.\ Phys.\ B {\bf 908} (2016) Pages 1-466 (July 2016),\\
	 {\tt http://www.sciencedirect.com/science/journal/05503213/908/supp/C}.
	 
	 \bibitem{XZbook}
	 Z.~Z.~Xing and S.~Zhou, {\it Neutrinos in Particle Physics, Astronomy and Cosmology} (Springer-Verlag, Berlin-Heidelberg, 2011).
	 
	 \bibitem{King:2013eh}
	 %
	 S.~F.~King,
	 J. Phys. G: Nucl. Part. Phys. {\bf 42} (2015) 123001
	 [arXiv:1510.02091].;
	 S.~F.~King and C.~Luhn,
	 Rept.\ Prog.\ Phys.\  {\bf 76} (2013) 056201
	 [arXiv:1301.1340];
	 
	 \bibitem{Abe:2017uxa}
	 K.~Abe {\it et al.} [T2K Collaboration],
	 Phys.\ Rev.\ Lett.\  {\bf 118} (2017) no.15,  151801
	 doi:10.1103/PhysRevLett.118.151801
	 [arXiv:1701.00432 [hep-ex]].
	 
	 \bibitem{Adamson:2017qqn} 
	 P.~Adamson {\it et al.} [NOvA Collaboration],
	 Phys.\ Rev.\ Lett.\  {\bf 118}, no. 15, 151802 (2017)
	 doi:10.1103/PhysRevLett.118.151802
	 [arXiv:1701.05891 [hep-ex]].
	 
	 \bibitem{NOvA_new}
	 ``Latest Oscillation Results from NOvA'',
	 Fermilab Joint Experimental-Theoretical Physics (JETP) seminar,
	 January 2018, NOVA-doc-25938-v3.
	 
	 \bibitem{deSalas:2017kay}
	 P.~F.~de Salas, D.~V.~Forero, C.~A.~Ternes, M.~Tortola and J.~W.~F.~Valle,
	 Phys.\ Lett.\ B {\bf 782} (2018) 633
	 doi:10.1016/j.physletb.2018.06.019
	 [arXiv:1708.01186 [hep-ph]].
	 
	 \bibitem{Esteban:2016qun} 
	 I.~Esteban, M.~C.~Gonzalez-Garcia, M.~Maltoni, I.~Martinez-Soler and T.~Schwetz,
	 JHEP {\bf 1701}, 087 (2017)
	 doi:10.1007/JHEP01(2017)087
	 [arXiv:1611.01514 [hep-ph]];
	 http://www.nu-fit.org/
	 
	 \bibitem{Minkowski:1977sc}
	 P. Minkowski, Phys. Lett. B {\bf 67}, 421 (1977).
	 
	 \bibitem{Yanagida:1979ss}
	 T. Yanagida, In {\it Proceedings of the Workshop on Unified Theory and
	 	the Baryon Number of the Universe}, edited by O. Sawada and A.
	 Sugamoto, (KEK, Tsukuba, 1979), p. 95.
	 
	 \bibitem{Gell-Mann:1979ss}
	 M. Gell-Mann, P. Ramond and R. Slansky, In {\it Supergravity},
	 edited by P. van Nieuwenhuizen and D. Z. Freeman, (North-Holland,
	 Amsterdam, 1979), p. 315.
	 
	 \bibitem{Glashow:1979ss}
	 S. L. Glashow, In {\it Quarks and Leptons}, edited by M. Levy {\it
	 	et al.} (Plenum, New York, 1980), p. 707.
	 
	 \bibitem{Mohapatra:1979ia}
	 R. N. Mohapatra and G. Senjanovic, Phys. Rev. Lett. {\bf 44}, 912 (1980).
	 
	 
\bibitem{Schechter:1980gr}
  J.~Schechter and J.~W.~F.~Valle,
  Phys.\ Rev.\ D {\bf 22} (1980) 2227;
  doi:10.1103/PhysRevD.22.2227
  J.~Schechter and J.~W.~F.~Valle,
  Phys.\ Rev.\ D {\bf 25} (1982) 774.
  doi:10.1103/PhysRevD.25.774
	 
	 \bibitem{King:1998jw}
	 S.~F.~King,
	 Phys.\ Lett.\ B {\bf 439} (1998) 350
	 doi:10.1016/S0370-2693(98)01055-7
	 [hep-ph/9806440];
	 S.~F.~King,
	 Nucl.\ Phys.\ B {\bf 562} (1999) 57
	 doi:10.1016/S0550-3213(99)00542-8
	 [hep-ph/9904210].
	 
	 \bibitem{King:1999mb}
	 S.~F.~King,
	 Nucl.\ Phys.\ B {\bf 576}, 85 (2000)
	 [hep-ph/9912492].
	 
	 \bibitem{King:2002nf}
	 S.~F.~King,
	 JHEP {\bf 0209}, 011 (2002)
	 [hep-ph/0204360].
	 
	 \bibitem{Frampton:2002qc}
	 P.~H.~Frampton, S.~L.~Glashow and T.~Yanagida,
	 Phys.\ Lett.\ B {\bf 548}, 119 (2002)
	 [hep-ph/0208157].
	 
	 \bibitem{Fukugita:1986hr}
	 M.~Fukugita and T.~Yanagida,
	 Phys.\ Lett.\ B {\bf 174}, 45 (1986).
	 
	 \bibitem{Harigaya:2012bw}
	 K.~Harigaya, M.~Ibe and T.~T.~Yanagida,
	 Phys.\ Rev.\ D {\bf 86}, 013002 (2012)
	 [arXiv:1205.2198].
	 
	 \bibitem{Zhang:2015tea}
	 J.~Zhang and S.~Zhou,
	 JHEP {\bf 1509}, 065 (2015)
	 [arXiv:1505.04858].
	 
	 \bibitem{Guo:2003cc}
	 W.~L.~Guo and Z.~Z.~Xing,
	 Phys.\ Lett.\ B {\bf 583}, 163 (2004)
	 [hep-ph/0310326].
	 
	 \bibitem{Ibarra:2003up}
	 A.~Ibarra and G.~G.~Ross,
	 Phys.\ Lett.\ B {\bf 591} (2004) 285
	 [hep-ph/0312138].
	 
	 \bibitem{Mei:2003gn}
	 J.~W.~Mei and Z.~Z.~Xing,
	 Phys.\ Rev.\ D {\bf 69}, 073003 (2004)
	 [hep-ph/0312167].
	 
	 \bibitem{Guo:2006qa}
	 W.~L.~Guo, Z.~Z.~Xing and S.~Zhou,
	 Int.\ J.\ Mod.\ Phys.\ E {\bf 16}, 1 (2007)
	 [hep-ph/0612033].
	 
	 \bibitem{Antusch:2011nz}
	 S.~Antusch, P.~Di Bari, D.~A.~Jones and S.~F.~King,
	 Phys.\ Rev.\ D {\bf 86} (2012) 023516
	 [arXiv:1107.6002].
	 
\bibitem{Branco:2005jr}
   G.~C.~Branco, M.~N.~Rebelo and J.~I.~Silva-Marcos,
   Phys.\ Lett.\ B {\bf 633} (2006) 345
   doi:10.1016/j.physletb.2005.11.067
   [hep-ph/0510412].
	 
	 \bibitem{King:2005bj}
	 S.~F.~King,
	 JHEP {\bf 0508} (2005) 105
	 doi:10.1088/1126-6708/2005/08/105
	 [hep-ph/0506297].
	 
	 \bibitem{Antusch:2011ic}
	 S.~Antusch, S.~F.~King, C.~Luhn and M.~Spinrath,
	 Nucl.\ Phys.\ B {\bf 856} (2012) 328
	 doi:10.1016/j.nuclphysb.2011.11.009
	 [arXiv:1108.4278 [hep-ph]].
	 
	 \bibitem{King:2013iva}
	 S.~F.~King,
	 JHEP {\bf 1307} (2013) 137
	 [arXiv:1304.6264].
	 
	 \bibitem{King:2015dvf}
	 S.~F.~King,
	 JHEP {\bf 1602}, 085 (2016)
	 [arXiv:1512.07531].
	 
	 \bibitem{King:2016yvg}
	 S.~F.~King and C.~Luhn,
	 JHEP {\bf 1609}, 023 (2016)
	 [arXiv:1607.05276].
	 
	 \bibitem{Ding:2018fyz}
	 G.~J.~Ding, S.~F.~King and C.~C.~Li,
	 arXiv:1807.07538 [hep-ph].
	 
	 \bibitem{King:2018kka}
	 S.~F.~King and C.~C.~Nishi,
	 arXiv:1807.00023 [hep-ph].
	 
	 \bibitem{Ballett:2016yod}
	 P.~Ballett, S.~F.~King, S.~Pascoli, N.~W.~Prouse and T.~Wang,
	 JHEP {\bf 1703} (2017) 110
	 doi:10.1007/JHEP03(2017)110
	 [arXiv:1612.01999 [hep-ph]].
	 
	 \bibitem{Bjorkeroth:2015tsa}
	 F.~Bj\"{o}rkeroth, F.~J.~de Anda, I.~de Medeiros Varzielas and S.~F.~King,	
  	 JHEP {\bf 1510} (2015) 104
  	 doi:10.1007/JHEP10(2015)104
  	 [arXiv:1505.05504 [hep-ph]].
	 
	 \bibitem{Bjorkeroth:2014vha}
	 F.~Bj\"{o}rkeroth and S.~F.~King,
	 J.\ Phys.\ G {\bf 42}, no. 12, 125002 (2015)
	 [arXiv:1412.6996].
	 
	 \bibitem{Bjorkeroth:2015ora}
	 F.~Bj\"orkeroth, F.~J.~de Anda, I.~de Medeiros Varzielas and S.~F.~King,
	 JHEP {\bf 1506} (2015) 141
	 [arXiv:1503.03306 [hep-ph]];
	 
	\bibitem{King:2013xba}
 	 S.~F.~King,
 	 Phys.\ Lett.\ B {\bf 724} (2013) 92
 	 doi:10.1016/j.physletb.2013.06.013
	  [arXiv:1305.4846 [hep-ph]].
  
	  \bibitem{King:2013hoa}
	  S.~F.~King,
	  JHEP {\bf 1401} (2014) 119
	  doi:10.1007/JHEP01(2014)119
	  [arXiv:1311.3295 [hep-ph]].
  
	  \bibitem{King:2014iia}
	   S.~F.~King,
	  JHEP {\bf 1408} (2014) 130
	  doi:10.1007/JHEP08(2014)130
	  [arXiv:1406.7005 [hep-ph]].
	 
	 \bibitem{Chianese:2018dsz}
	 M.~Chianese and S.~F.~King,
	 arXiv:1806.10606 [hep-ph].
		 
	 \bibitem{King:2016yef}
	 S.~F.~King, J.~Zhang and S.~Zhou,
	 JHEP {\bf 1612} (2016) 023
	 doi:10.1007/JHEP12(2016)023
	 [arXiv:1609.09402 [hep-ph]].
		 
	 \bibitem{Geib:2017bsw}
	 T.~Geib and S.~F.~King,
	 Phys.\ Rev.\ D {\bf 97} (2018) no.7,  075010
	 doi:10.1103/PhysRevD.97.075010
	 [arXiv:1709.07425 [hep-ph]].
	 
	 \bibitem{Antusch:2005gp}
	 S.~Antusch, J.~Kersten, M.~Lindner, M.~Ratz and M.~A.~Schmidt,
	 JHEP {\bf 0503} (2005) 024
	 doi:10.1088/1126-6708/2005/03/024
	 [hep-ph/0501272].
	 
	 \bibitem{Weinberg:1979}
	 S.~Weinberg,
	 Phys. Rev. Lett. {\bf 42} (1979) 850-853
	 HUTP-78/A040.
	 
	 \bibitem{Toussaint:1979}
	 D.~Toussaint, S.~B.~Treiman, Frank Wilczek, A.~Zee,
	 Phys. Rev. D {\bf19} (1979) 1036-1045
	 
	 \bibitem{Ade:2015xua}
	 P.~A.~R.~Ade {\it et al.} [Planck Collaboration],
	 Astron.\ Astrophys.\  {\bf 594} (2016) A13
	 doi:10.1051/0004-6361/201525830
	 [arXiv:1502.01589 [astro-ph.CO]].
	 
	 \bibitem{Sakharov:1967}
	 A.~D.~Sakharov,
	 Pisma Zh. Eksp. Teor. Fiz. {\bf 5}, 32 (1967)
	 JETP Lett. {\bf 5}, 24 (1967)
	 
	 \bibitem{Buchmuller:2004}
	 W.~Buchmuller, P.~Di Bari, M.~Plumacher,
	 Annals Phys. {\bf 315} (2005) 305-351
	 
	 \bibitem{Abada:2006ea}
	 A.~Abada, S.~Davidson, A.~Ibarra, F.-X.~Josse-Michaux, M.~Losada and A.~Riotto,
	 JHEP {\bf 0609} (2006) 010
	 doi:10.1088/1126-6708/2006/09/010
	 [hep-ph/0605281].
	 
	 \bibitem{Antusch:2006}
	 S.~Antusch, S.~F.~King, A.~Riotto,
	 JCAP {\bf 0611} (2006) 011
	 [hep-ph/0609038]
	 
	 \bibitem{Covi:1996}
	 L.~Covi, E.~Roulet, F.~Vissani,
	 Phys. Lett. B {\bf 384} (1996) 169-174
	 [hep-ph/9605319]
	 
	 \bibitem{Blanchet:2007}
	 S.~Blanchet, P.~Di Bari
	 JCAP {\bf 0703} (2007) 018
	 [hep-ph/0607330]
	 
	\bibitem{Dev:2017trv}
	P.~S.~B.~Dev, P.~Di Bari, B.~Garbrecht, S.~Lavignac, P.~Millington and D.~Teresi,
	Int.\ J.\ Mod.\ Phys.\ A {\bf 33} (2018) 1842001
	doi:10.1142/S0217751X18420010
	[arXiv:1711.02861 [hep-ph]].
	 
	 \bibitem{Blanchet:2006}
	 S.~Blanchet, P.~Di Bari
	 JCAP {\bf 0606} (2006) 023
	 [hep-ph/0603107]
	 
	 \bibitem{King:2002}
	 S.~F.~King
	 Phys. Rev. D {\bf 67} (2003) 113010
	 [hep-ph/0211228]
	
\end{thebibliography}
\end{document}